%% file: samplepaper.tex
\documentclass[runningheads]{llncs}
\usepackage[T1]{fontenc}
\usepackage{graphicx}

\usepackage[linesnumbered, ruled, vlined]{algorithm2e}
\usepackage{longtable}
\usepackage{array}
\usepackage{subfigure}
\usepackage{stfloats}
\usepackage{balance}
\usepackage{multicol}
\usepackage{color}
\usepackage{bm}
\usepackage{booktabs} % For formal tables
\usepackage{epsfig}
\usepackage{enumitem}
\usepackage{balance}
\usepackage{amssymb}
\usepackage{multirow}

\usepackage{arydshln}

\usepackage{amsmath}
\usepackage{cleveref}
\usepackage{graphicx}
\usepackage{textcomp}
\usepackage{xcolor}
\usepackage{subfigure}

\usepackage{tabularx}
\usepackage{bbm}

\newcommand{\hide}[1]{} %hide
\newcommand{\vpara}[1]{\vspace{0.1in}\noindent\textbf{#1}}

\newcommand{\secref}[1]{Section~\ref{#1}} %section reference
\newcommand{\beqn}[1]{\vspace{-0.03in}\begin{eqnarray}#1\end{eqnarray}\vspace{-0.03in}}

\newcommand{\RC}{${\rm SGG}$}
\newcommand{\sRC}{${\rm SGG}$\space}
\newcommand{\tRC}{$\textbf{SGG}$}
\newcommand{\stRC}{$\textbf{SGG}$\space}
\begin{document}
\title{Self-supervised Graph Learning for Occasional Group Recommendation}
\author{Bowen Hao\inst{1} \and
Hongzhi Yin\thanks{Corresponding Author} \inst{2}\and Cuiping Li\inst{3} \and Hong Chen\inst{3}}
\authorrunning{B. Hao et al.}
\institute{Capital Normal University, China \and
The University of Queensland, Australia \and Renmin University of China, China\\
\email{haobowenmail@163.com,  h.yin1@uq.edu.au,  \{licuiping,chong\}@ruc.edu.cn}}
\maketitle              % typeset the header of the contribution
\begin{abstract}
	As an important branch in Recommender System, occasional group recommendation has received more and more attention. In this scenario, each occasional group (cold-start group) has no or few historical interacted items. As each occasional group has extremely sparse interactions with items, traditional group recommendation methods can not learn high-quality group representations.
	The recent proposed Graph Neural Networks (GNNs), which incorporate the high-order neighbors of the target occasional group, can alleviate the above problem in some extent.  However, these GNNs still can not explicitly strengthen the embedding quality of the high-order neighbors with few interactions. Motivated by the Self-supervised Learning technique, which is able to find the correlations within the data itself, we propose a self-supervised graph learning framework, which takes the user/item/group embedding reconstruction as the pretext task to enhance the embeddings of the cold-start users/items/groups.
    In order to explicitly enhance the high-order cold-start neighbors‘ embedding quality, we further introduce an embedding enhancer, which leverages the self-attention mechanism to improve the embedding quality for them. Comprehensive experiments show the advantages of our proposed framework than the state-of-the-art methods.

\keywords{Occasional group recommendation \and Self-supervised learning \and Graph neural network}
\end{abstract}

\input{intro.tex}

\input{problem.tex}

\input{approach.tex}

\input{exp.tex}

\input{related.tex}

\input{conclusion.tex}


\end{document}

%% file: intro.tex
\section{Introduction}
\label{sec:intro}

\begin{figure}[t]
	\centering
	\includegraphics[width= 0.90 \textwidth]{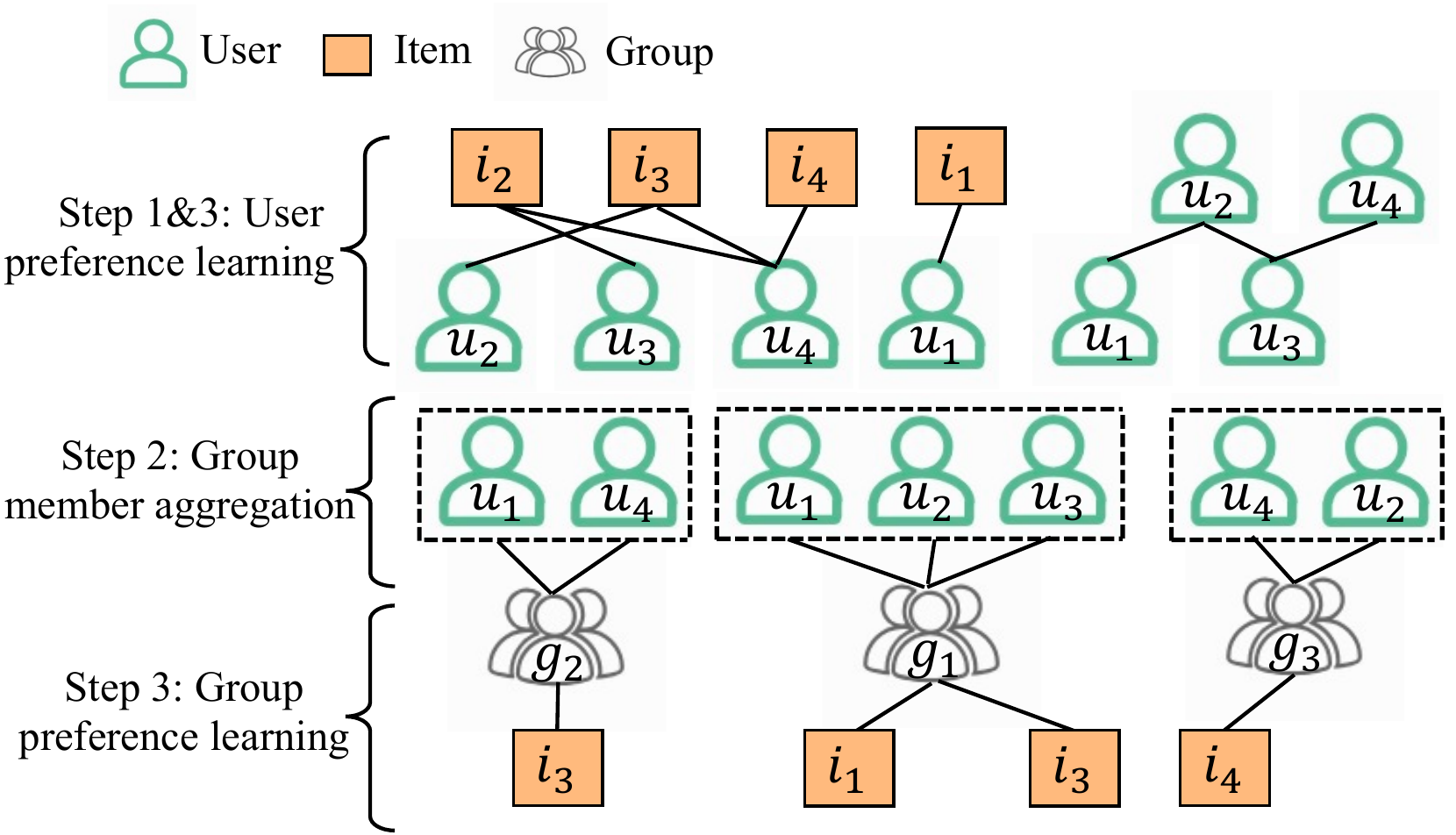}
	\caption{\label{fig:GNN_group_recommendation} A running example for group recommendation.}
\end{figure}

As an important branch in Recommender System~\cite{journal1_citation,journal2_citation,journal3_citation}, occasional group recommendation has received more and more attention, and many social media platforms such as Meetup and Facebook aims to solve this problem~\cite{aaai/LiuWL0SH19,sigir/SunW018,cikm/GaoWQZYH16,icde/YinZNHZ18}.
This task can be formulated as recommending items to occasional groups, where each occasional group has no or few historical interacted items\footnote{This paper addresses the occasional group (a.k.a cold-start group) with few historical interacted items.}. Since each occasional group has extremely sparse interactions with items, traditional group recommendation methods can not learn high-quality group representations.

To solve this problem,
some early studies adopt heuristic pre-defined aggregation strategy such as average strategy~\cite{BaltAverage}, least misery strategy~\cite{pvldb/Amer-YahiaRCDY09} and maximum satisfaction strategy~\cite{BorattoC11} to aggregate the user preference to obtain the group preference. However, due to the fixed aggregation strategies, 
These methods are difficult to capture the complex dynamic process of group decision making, which results in
unstable recommendation performance~\cite{mta/PessemierDM14}.
Further, Cao et al.~\cite{sigir/Cao0MAYH18} propose to assign each user an attention weight, which represents the  influence of group member in deciding the group's choice on the target item.
However, when some users in the occasional group only interact with few items (a.k.a. cold-start users),  the attention weight assigned for each user is diluted by these cold-start users, and thus results in biased group profile.

Recently, a few Graph Neural Networks (GNNs) based group recommendation methods are proposed~\cite{sigir/SankarWWZYS20,icde/GuoYW0HC20,sigir/Wang0RQ0LZ20,tois/GUOYintois21group}. The core idea of these GNNs is to incorporate high-order neighbors as collaborative signals to strengthen the  cold-start users' embedding quality, and further enhance the embedding quality of the target group.
As shown in Figure~\ref{fig:GNN_group_recommendation}, the GNN model first conducts graph convolution multiple steps on the user-item and user-user interaction graphs to learn the preference of group members, and then performs average~\cite{tois/GUOYintois21group}, summation and pooling~\cite{conf/nips/ZaheerKRPSS17} or attention mechanism~\cite{tois/GUOYintois21group}  to aggregate  the preferences of group members to obtain the group representation. 
Next, based on the aggregated embeddings of groups and users, the GNNs can estimate the probability that a group/user purchase an item.  
Finally, the recommendation-oriented loss (e.g., BPR loss~\cite{uai/RendleFGS09}) is used to optimize the groups/users/items embeddings.
Moreover, Zhang et al.~\cite{conf/CIKM21/JMJLJH21} propose hypergraph convolution network (HHGR) with self-supervised node dropout strategy, which can model complex high-order interactions between groups and users. Through incorporating self-supervised signals, HHGR can alleviate the cold-start issue in some extent.

However, the above methods still suffer from the following challenges. 
First, the group representation not only depends on the group members' preferences, but also relies on the group-level preferences towards items and collaborative group signals (the groups that  share common users/items).
Although some GNNs consider either group-level  preferences~\cite{sigir/HeCZ20} or collaborative group signals~\cite{tois/GUOYintois21group,conf/CIKM21/JMJLJH21} to form the group representation, they do not consider all these signals together.
Second, the GNNs can not explicitly enhance the high-order cold-start neighbors' embedding quality.
For example in Figure~\ref{fig:GNN_group_recommendation}, for the target group $g_2$, its group member $u_1$ and high-order neighbor $i_1$ only have few interactions. The embeddings of $u_1$ and $i_1$ are inaccurate, which will hurt the embedding quality of $g_2$ after performing graph convolution operation.
Thus, \textit{how can we learn high-quality embeddings by GNNs for occasional group recommendation?}

\hide{Fourth, in some cases  when users form a group, they may pursue a target
that is different from the preference of each user. We believe one possible reason is that choosing a target item is the result of group members compromising their interests. For example in Fig.~\ref{fig:GNN_group_recommendation}, a child $u_1$ prefers cartoon movie $i_2$ and her parents $u_2$ and $u_3$
like romantic movie $i_3$ and $i_4$; but when they go to a cinema together, the final chosen movie could be an educational movie $i_1$, as the parents may think it is crucial to educate their children.

 This leads us the following two research questions: \textit{1) how can we learn more accurate group/user embeddings by GNNs?}   \textit{2) How can we model the group compromise decision process?  }
}

To this end,
motivated by the self-supervised learning (SSL) technique~\cite{kdd/HuDWCS20,kdd/QiuCDZYDWT20,hao2021pre}, which aims to spontaneously find the supervised signals from the input data itself and can further benefit the downstream tasks~\cite{TKDE_citation1,TKDE_citation2}, we propose a self-supervised graph learning framework (\RC), which reconstructs the user/item group embedding with the backbone GNNs from multiple interaction graphs under the meta-learning setting~\cite{nips/VinyalsBLKW16}.
\sRC can explicitly improve the users'/items'/groups' embedding quality.
More concretely, first, we choose the groups/users/items that have enough interactions as the target groups/users/items. We treat the learned embeddings of these target groups/users/items as the ground-truth embeddings, since traditional recommendation algorithms can learn high-quality embeddings for them.
Further, we mask a large proportion neighbors of the target  groups/users/items to simulate the cold-start scenario, where the cold-start groups/users/items have few interactions. Based on the partially observed neighbors of the target groups/users/items, we perform graph convolution multiple steps  to learn their ground-truth embeddings. More concretely, for each target group/user/item in their counterpart interaction graphs\footnote{For each group, its counterpart graphs are group-group, group-item and group-user graphs; for each user, his counterpart graphs are user-user and user-item graphs; for each item, its counterpart graph is user-item graph.}, we randomly sample $K$ first-order neighbors for them. Based on the sampled neighbors, we perform graph convolution operation multiple times in each interaction graph, and fuse the corresponding refined embeddings to predict the target embedding. Finally, we jointly optimize the reconstruction losses between the users'/items'/groups' ground-truth embeddings and the predicted embeddings by the GNNs, which improves the convolution ability of the GNNs. Thus, the GNNs can obtain high-quality group representation that contains all the signals as presented in the first challenge.

Nevertheless, the above proposed embedding reconstruction task does not explicitly strengthen the high-order cold-start neighbors' embedding quality  for the target users/items/groups. To explicitly enhance the high-order cold-start neighbors' embedding quality, we incorporate an embedding enhancer to learn high-quality node's embeddings under the same meta-learning setting as mentioned before. The embedding enhancer is instantiated as the self-attention learner, which learns the cold-start nodes' embeddings based on their masked first-order neighbors in the counterpart interaction graphs. 
We incorporate the meta embedding produced by the embedding enhancer into each graph convolution step to further  strengthen the  GNNs' aggregation ability.  
The contributions are:
\hide{We further propose multi-armed bandit mechanism to balance the exploration and exploitation problem.}

\begin{itemize}[ leftmargin=10pt ]
	\item We present a self-supervised graph learning framework for group recommendation.  In this framework, we design the user/item/group embedding reconstruction task with GNNs under the meta-learning setting, 

	\item We further introduce an embedding enhancer  to strengthen the  GNNs' aggregation ability, which can improve the high-order cold-start neighbors' embedding quality.

	\item Comprehensive experiments show the superiority of our proposed framework against the state-of-the-art methods.
	
	\hide{Compared with the original GNN models, the pre-training GNN models achieve 15.3-53.4\% improvement in terms of Spearman correlation~\cite{ziniuhufewshot19} between the predicted embeddings and the ground truth embeddings by the intrinsic evaluation on inferring user/item embeddings, and also achieve  6.9-13.2\% improvement in NDCG@20 by the extrinsic evaluation on the downstream recommendation task.}
	
\end{itemize}

%% file: problem.tex
\section{Preliminary}

\subsection{Problem Definition}
There are three sets of entities in the group recommendation scenario: $U = \ \{  u_1, \cdots, u_{|U|}  \}$ denotes a user set, $I = \ \{  i_1, \cdots, i_{|I|} \}$ denotes an item set, and $G=\  \{  g_1, \cdots, g_{|G|} \}$ denotes a group set.  There are three kinds of observed interaction graphs, i.e., group-item   subgraph  $\mathcal{G}_{GI}$, user-item  subgraph  $\mathcal{G}_{UI}$ and  group-user subgraph $\mathcal{G}_{GU}$.
Since the social connections of user-user and group-group are also important to depict the user and group profiles, we build two kinds of implicit interaction graphs based on $\mathcal{G}_{GI}$ and $\mathcal{G}_{UI}$, namely user-user subgraph $\mathcal{G}_{UU}$ and group-group subgraph $\mathcal{G}_{GG}$. In $\mathcal{G}_{UU}$, the two users are connected if they share with more than $c_u$ items.
Similarly, in $\mathcal{G}_{GG}$, the two groups are connected if they share with more than $c_g$ items.
Formally, notation $\mathcal{G}$ = $\{  \mathcal{V}, \mathcal{E}  \}$ denotes the set of observed and implicit interaction graphs, i.e.,  $\mathcal{G}$ = $\mathcal{G}_{GI}  \cup \mathcal{G}_{UI} \cup \mathcal{G}_{GU} \cup \mathcal{G}_{GG} \cup \mathcal{G}_{UU}$, where $\mathcal{E}$ is the edge set and $\mathcal{V}$ is the nodes set which contains $\{ U, I, G\} $.
%Based on these three sets, we introduce the following definitions.

\hide{
\vpara{Definition 1.} \textit{Interaction Graph.}  An interaction graph describes heterogeneous connections among groups, users and items in group recommendation scenario. 
Formally, it is denoted as a heterogeneous graph $\mathcal{G} = \{ \mathcal{V}, \mathcal{E} \}$,  where $\mathcal{V}$ is the set
of nodes $\{ U, I, G\} $, and $\mathcal{E}$ is the set of edges $\{\mathcal{E}_{UG}, \mathcal{E}_{UI}, \mathcal{E}_{GI}, \mathcal{E}_{GG}, \mathcal{E}_{UU} \}$. Notation $\mathcal{E}_{UG} $ is the subset of edges between users and groups, $\mathcal{E}_{GI}$
is the subset of edges between groups and items,  $\mathcal{E}_{UI}$ is the
subset of edges between users and items, $\mathcal{E}_{UU}$ is the
subset of edges between users and $\mathcal{E}_{GG}$ is the
subset of edges between groups.
The whole interaction graph $\mathcal{G}$ can be divided into  five  subgraphs, i.e., user-group  subgraph  $\mathcal{G}_{UG} = \{ U \cup G, \mathcal{E}_{UG} \}$, group-item   subgraph  $\mathcal{G}_{GI} = \{ G \cup I, \mathcal{E}_{GI} \}$, user-item  subgraph  $\mathcal{G}_{UI} = \{ U \cup I, \mathcal{E}_{UI} \}$, user-user  subgraph $\mathcal{G}_{UU} = \{ U, \mathcal{E}_{UU} \}$ and group-group subgraph $\mathcal{G}_{GG} = \{ G, \mathcal{E}_{GG} \}$.
}
%\vpara{Definition 2.} \textit{GNN for Group Recommendation.}  Given the interaction graph $\mathcal{G}$, we aim to train the GNN-based encoder $f$ that can recommend top-$k$ items for the target group $g$.

\vpara{Definition 1.} \textit{GNN-oriented Group Recommendation.}  Given the interaction graph $\mathcal{G}$, the goal is to train a GNN-based encoder $f$ that can recommend top-$k$ items for the target group $g$.

\subsection{Base GNN for Group Recommendation}
\label{sec:base_gnn}
Although existing GNN-based group recommendation methods show their diversity in modelling group interactions with users
and items~\cite{sigir/Wang0RQ0LZ20,icde/GuoYW0HC20,sigir/HeCZ20,tois/GUOYintois21group}, we notice that they essentially share a general model structure. Based on this finding, we present a base GNN model, which consists of a representation learning module  and a jointly training module. 
The representation learning module learns the representations of groups and users upon their counterpart interaction graphs, while the jointly training module  optimizes the user/group preferences over items to compare the likelihood and  the true user-item/group-item observations.
%Unifying these GNN-based group recommender within a single framework facilitates deeper analysis into their shortcomings in addressing occasional groups.

\hide{
	\beqn{
		\label{eq:basic_graph_convolution_u}
		\textbf{h}_{u_{UI}}^{l} &=&  \text{CONV}( \textbf{h}_{u_{UI}}^{l-1}  ,   \textbf{h}^{l}_{  \mathcal{N}(u_{UI})}  ), \\ \nonumber
		\textbf{h}_{u_{UU}}^{l} &=& \text{CONV}( \textbf{h}_{u_{UU}}^{l-1}  ,   \textbf{h}^{l}_{  \mathcal{N}(u_{UU})}  ),	\\ \nonumber
}}

\hide{
	\beqn{
		\label{eq:aggregate_u}
		\textbf{h}_u^{L} &=& \sum_{c \in \{ UI, UU  \}} a_{c} 	\textbf{h}_{u_{c}}^{L},  \\ \nonumber
		a_c &=& \frac{ {\rm exp} (\mathbf{W}_c 	\textbf{h}_{u_{c}}^{L} ) }{\sum_{c' \in \{  UI, UU\}}  {\rm exp} (\mathbf{W}_{c'} 	\textbf{h}_{u_{c'}}^{L} )}, 
	}
}

\subsubsection{Representation Learning Module.}  This module first learns the user representation upon the user-item and user-user subgraphs, and then learns the group representation upon the group-group, group-item and group-user subgraphs.
Specifically, for each user $u$, we first sample his first-order neighbors on   $\mathcal{G}_{UI}$ and  $\mathcal{G}_{UU}$, and then perform graph convolution:
$ \textbf{h}_{u_{UI}}^{l}=  \text{CONV}( \textbf{h}_{u_{UI}}^{l-1}  ,   \textbf{h}^{l}_{  \mathcal{N}(u_{UI})}  )  $, $\textbf{h}_{u_{UU}}^{l} = \text{CONV}( \textbf{h}_{u_{UU}}^{l-1}  ,   \textbf{h}^{l}_{  \mathcal{N}(u_{UU})}  )$, where $\text{CONV}$ can be instantiated into any GNN models, such as LightGCN~\cite{sigir/0001DWLZ020} or GCN~\cite{Thomasgcn}.  $\textbf{h}_{u_{UI}}^{l}$ and $\textbf{h}_{u_{UU}}^{l}$   denote the user embeddings calculated from  $\mathcal{G}_{UI}$ and  $\mathcal{G}_{UU}$ at the $l$-th graph convolution step, $\textbf{h}_{u_{UI}}^{0}$ and  $\textbf{h}_{u_{UU}}^{0}$  are randomly initialized embeddings. $ \textbf{h}^{l}_{  \mathcal{N}(u_{UI})} $ and $ \textbf{h}^{l}_{  \mathcal{N}(u_{UU})}$  mean the averaged neighbor embeddings, where the neighbors are sampled from $\mathcal{G}_{UI}$ and  $\mathcal{G}_{UU}$, respectively. After performing $L$-th convolution steps, we can obtain the refined user embeddings $\textbf{h}_{u_{UI}}^{L}$ and $\textbf{h}_{u_{UU}}^{L}$  from the counterpart  subgraphs.  Finally we conduct the soft-attention algorithm~\cite{tkde/HeHSLJC18} to aggregate these embeddings to form the final user embedding $\textbf{h}_{u}^{L}$:
$	\textbf{h}_u^{L} = \sum_{c \in \{ UI, UU  \}} a_{c} 	\textbf{h}_{u_{c}}^{L}$, $	a_c = \frac{ {\rm exp} (\mathbf{W}_c 	\textbf{h}_{u_{c}}^{L} ) }{\sum_{c' \in \{  UI, UU\}}  {\rm exp} (\mathbf{W}_{c'} 	\textbf{h}_{u_{c'}}^{L} )}$, where $\{ \mathbf{W}_c | c \in \{  UI, UU \}\}$ are trainable parameters, $\{  a_c | c \in \{   UI, UU   \}$ are the learned attention weight for each subgraph.

Then for each group $g$, we first sample its first-order neighbors
from the $\mathcal{G}_{GI}$,  $\mathcal{G}_{GU}$ and $\mathcal{G}_{GG}$,  and perform graph convolution:
\beqn{
	\label{eq:basic_graph_convolution}
	\textbf{h}_{g_{GI}}^{l} &=&  \text{CONV}( \textbf{h}_{g_{GI}}^{l-1}  ,   \textbf{h}^{l}_{  \mathcal{N}(g_{GI})}  ), \\ \nonumber
	\textbf{h}_{g_{GU}}^{l} &=& \text{CONV}( \textbf{h}_{g_{GU}}^{l-1}  ,   \textbf{h}^{l}_{  \mathcal{N}(g_{GU})}  ),	\\ \nonumber
	\textbf{h}_{g_{GG}}^{l} &=& \text{CONV}( \textbf{h}_{g_{GG}}^{l-1}  ,   \textbf{h}^{l}_{  \mathcal{N}(g_{GG})}  ), \\ \nonumber
}

\noindent where  $\textbf{h}_{g_{GI}}^{l}$, $\textbf{h}_{g_{GU}}^{l}$ and $\textbf{h}_{g_{GG}}^{l}$  denote the group embeddings calculated from  $\mathcal{G}_{GI}$,  $\mathcal{G}_{GU}$ and $\mathcal{G}_{GG}$ at the $l$-th graph convolution step, $\textbf{h}_{g_{GI}}^{0}$, $\textbf{h}_{g_{GU}}^{0}$ and $\textbf{h}_{g_{GG}}^{0}$ are randomly initialized embeddings. $ \textbf{h}^{l}_{  \mathcal{N}(g_{GI})} $, $ \textbf{h}^{l}_{  \mathcal{N}(g_{GU})}$ and $ \textbf{h}^{l}_{  \mathcal{N}(g_{GG})}$ mean the averaged neighbor embeddings, where the neighbors are sampled from $\mathcal{G}_{GI}$,  $\mathcal{G}_{GU}$ and $\mathcal{G}_{GG}$, respectively. After performing $L$-th convolution steps, we can obtain the refined group embeddings $\textbf{h}_{g_{GI}}^{L}$, $\textbf{h}_{g_{GU}}^{L}$ and $\textbf{h}_{g_{GG}}^{L}$ from these three subgraphs. 
Same as existing works~\cite{sigir/HeCZ20,sigir/HeCZ20,tois/GUOYintois21group,conf/CIKM21/JMJLJH21}, we further  average the first-order neighbors in $\mathcal{G}_{GU}$ to obtain the aggregated group embedding $\textbf{h}_{g_{GU'}}^{L}$,
\beqn{
\label{eq:fuse}
\textbf{h}_{g_{GU'}}^{L} = f_{agg} ( \{ \textbf{h}_{u_{GU}}^{L} |  u \in \mathcal{N}(g_{GU}) \} ),
}

\noindent where $\textbf{h}_{u_{GU}}^{L}$ is obtained  by performing graph convolution $L$ steps in $\mathcal{G}_{GU}$, $\mathcal{N}(g_{GU}) $ denotes the  first-order user set sampled from $\mathcal{G}_{GU}$, $f_{agg}$ is the aggregate function such as  average~\cite{tois/GUOYintois21group}, summation and pooling~\cite{conf/nips/ZaheerKRPSS17}, or self-attention mechanism~\cite{nips/VaswaniSPUJGKP17}. 
In our experiments, we find  attention mechanism has the best performance.
Finally, we use the soft-attention algorithm to aggregate the above embeddings  to form the final group embedding $\textbf{h}_{g}^{L}$:
\beqn{
	\label{eq:aggregate}
	\textbf{h}_g^{L} &=& \sum_{c \in \{ GI, GU, GU', GG  \}} a_{c} 	\textbf{h}_{g_{c}}^{L},  \\ \nonumber
	a_c &=& \frac{ {\rm exp} (\mathbf{W}_c 	\textbf{h}_{g_{c}}^{L} ) }{\sum_{c' \in \{  GI, GU, GU', GG \}}  {\rm exp}  (\mathbf{W}_{c'} 	\textbf{h}_{g_{c'}}^{L} )}, \\ \nonumber
}

\noindent where $\{ \mathbf{W}_c | c \in \{  GI, GU, GU', GG \}\}$ are trainable parameters, $\{  a_c \ | \ c \in \{   GI,  GU, \\ GU',  GG  \}$ are the learned attention weights.

\subsubsection{Jointly Training Module.}  This module jointly optimizes the user preferences over items with the user-item loss $\mathcal{L}_u$ and the group preferences over items with the group-item loss $\mathcal{L}_g$, i.e.,  $\mathcal{L}_{main} = 	\mathcal{L}_{g} + \lambda 	\mathcal{L}_{u} $, where  $\mathcal{L}_{main}$ is the final recommendation loss with a balancing hyper-parameter $\lambda$.  Here, we use BPR loss~\cite{uai/RendleFGS09} to calculate $\mathcal{L}_u$ and $\mathcal{L}_g$:
 \beqn{
	\label{eq:bprloss}
	\mathcal{L}_{u} &=& \sum_{ (u,i)\in \mathcal{E}_{UI}, (u,j)\notin \mathcal{E}_{UI} }  - \ln \sigma( y(u, i) - y(u, j) ) ,  \\ \nonumber
	\mathcal{L}_{g} &=& \sum_{ (g,i)\in \mathcal{E}_{GI}, (g,j)\notin \mathcal{E}_{GI} }  - \ln \sigma( y(g, i) - y(g, j) ) ,  \\ \nonumber 
}

\noindent  where $\sigma$ is the activation function, $y(u, i) = {\textbf{h}_{u}^L}^\mathrm{T} \textbf{h}_{i}^L$, $y(g, i) = {\textbf{h}_{g}^L}^\mathrm{T} \textbf{h}_{i}^L$,  $\mathcal{E}_{UI}$ and $\mathcal{E}_{GI}$ represent the edges in $\mathcal{G}_{UI}$ and $\mathcal{G}_{GI}$.

Although the above presented GNNs can address the occasional groups through
incorporating  high-order collaborative signals, they still can not deal with the groups/users/items with few interactions, and thus can not learn high-quality embeddings for them.

%% file: approach.tex
\section{The proposed model}
\begin{figure}[t]
	\centering
	\includegraphics[width= 0.90 \textwidth]{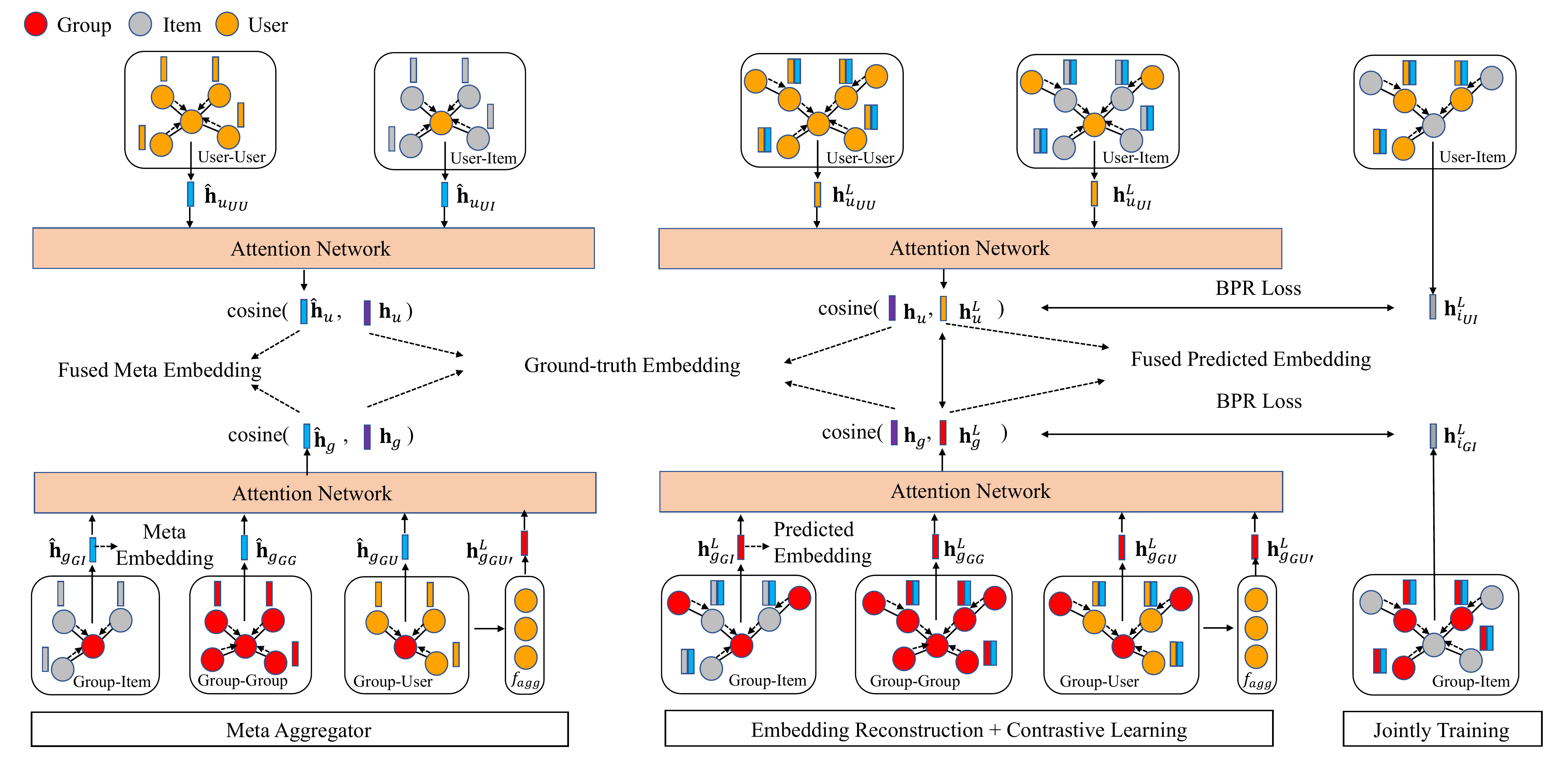}
	\caption{\label{fig:overall_framework} The overall framework of \tRC. \small{\stRC contains a self-attention based embedding enhancer,  which  incorporates the generated meta embedding at  each step of the original GNN convolution process.} }
\end{figure}

We propose a  self-supervised graph learning framework for group recommendation (\RC).
We first describe the process of embedding reconstruction with GNN, and then detail an embedding enhancer which is incorporated into the backbone GNN model to further enhance the embedding quality.
Finally, we present how \sRC is trained and analyze its time complexity.
The overall framework of \sRC is shown in Figure~\ref{fig:overall_framework}.

\subsection{ Embedding Reconstruction with GNN}
\label{sec:basic_reconstruction}
We propose  embedding reconstruction with GNN, which jointly reconstructs the groups/users/items embeddings from multiple subgraphs under the meta-learning setting.
Here we take the group embedding reconstruction as an example. Notably, the user/item embedding reconstruction process is similar to the group embedding reconstruction process.
Specifically, first, we need the groups with abundant interactions as the target groups, and use any recommendation model such as AGREE~\cite{sigir/Cao0MAYH18} or LightGCN~\cite{sigir/0001DWLZ020} to learn the embeddings of the target groups as the ground-truth embeddings\footnote{Previous paper~\cite{hao2021pre} has demonstrated that such recommendation models can obtain  high-quality embeddings of nodes with enough interactions.}. Then we mask a large proportion neighbors of the target group to simulate the occasional group. Based on the remained neighbors, we repeat the graph convolution operation multiple times to learn the ground-truth embeddings. 
Formally, for each target group $g$, we use notation $\textbf{h}_g$ to represent its ground-truth embedding.
To mimic the occasional
group, in each training episode, for each target group, we randomly sample $K$ items, $K$ users and $K$ groups from the corresponding group-item subgraph $\mathcal{G}_{GI}$, group-user subgraph $\mathcal{G}_{GU}$ and group-group subgraph $\mathcal{G}_{GG}$. 
We sample neighbors $L$ steps for each subgraph, i.e., for each target group, in each subgraph, we sample  from its first-order neighbors to its $L$-order neighbors. After the sampling process is finished, for each target group,  and for each subgraph, we can obtain at most $K^l (1 \leq l \leq L)$ $l$-order neighbors. 
Next we use Eq.~\eqref{eq:basic_graph_convolution} to conduct the graph convolution operation $L$ steps from scratch to obtain the refined group embeddings $\textbf{h}_{g_{GI}}^{L}$, $\textbf{h}_{g_{GU}}^{L}$ and $\textbf{h}_{g_{GG}}^{L}$,  use Eq.~\eqref{eq:fuse} to obtain the aggregated group embedding $\textbf{h}_{g_{GU'}}^{L}$, and use Eq.~\eqref{eq:aggregate}  to obtain the fused group embedding $\textbf{h}_{g}^{L}$.
Finally, following~\cite{conf/acl/HuCCS19,hao2021pre}, we measure the cosine similarity between $\textbf{h}_g$ and $\textbf{h}_{g}^{L}$, since the cosine similarity is a popularity indicator for the semantic similarity between embeddings:

\beqn{
	\label{eq:consine_similarity}
	\mathcal{L}_{R_g}:  \mathop{\arg\max}_{\Theta_f} \sum_{g} {\rm cos}( \textbf{h}_g^{L}, \textbf{h}_g ),
}

\noindent where $\Theta_f $ is the parameters of the GNN model $f$.  Similarly, we reconstruct the user embedding based on  $\mathcal{G}_{UI}$ and $\mathcal{G}_{UU}$ with loss $\mathcal{L}_{R_u}$, and reconstruct the item embedding based on $\mathcal{G}_{UI}$ with loss $\mathcal{L}_{R_i}$.
In practice, we jointly optimize group/user/item embedding reconstruction tasks with loss $\mathcal{L}_{R}$:

\beqn{
	\label{eq:over_all_reconstruction}
	\mathcal{L}_{R} = 	\mathcal{L}_{R_g} + 	\mathcal{L}_{R_u} + 	\mathcal{L}_{R_i}   .
}
\quad Notably, the above proposed embedding reconstruction task is trained under the meta-learning setting, which enables the GNNs rapidly being adapted to new occasional groups. After we trained the model, when it comes to a new occasional group, given its first- and high-order neighbors,
the pre-trained GNNs can generate a more accurate embedding for it.
However, the above proposed embedding reconstruction task does not 
explicitly strengthen the high-order cold-start neighbors' embedding quality. 
For example, if the high-order cold-start neighbors' embeddings are biased, they will affect the embeddings of the target group when performing graph convolution. As shown in Fig.~\ref{fig:GNN_group_recommendation}, for the target group $g_2$, its group member $u_1$ and high-order neighbor $i_1$ only have few interactions. The embeddings of $u_1$ and $i_1$ are inaccurate, which will hurt the embedding quality of $g_2$ after performing graph convolution operation.
%Although some classical GNN models such as GrageSAGE~\cite{nips/HamiltonYL17} or FastGCN~\cite{iclr/ChenMX18} filter neighbors before aggregating them, they usually follow the random or importance sampling strategies, which still ignores the cold-start characteristics of the high-order cold-start neighbors. 
To solve this problem, we further incorporate an embedding enhancer  into the above embedding reconstruction GNN model.

\subsection{Embedding  Enhancer}

\label{sec:meta_reconstruction}
To explicitly strengthen the embedding quality of the high-order cold-start neighbors,  we propose the embedding enhancer, which also learns the ground-truth group/user/item embedding, but only based on the first-order neighbors of the target group/user/item sampled from the counterpart graphs.
Specifically, before we train the above embedding reconstruction task with GNNs, we train the embedding enhancer $f_{meta}$ under the same meta-learning setting as proposed in~\secref{sec:basic_reconstruction}.
Once the embedding enhancer $f_{meta}$ is trained, we combine the enhanced group/user/item embedding\footnote{The enhanced group/user/item embedding is an additional embedding produced by $f_{meta}$.} produced by $f_{mata}$ with the original group/user/item embedding at each graph convolution step to improve the cold-start neighbors' embeddings.
Notably,  the GNN model is to strengthen the cold-start groups'/users'/items' embedding quality, while the embedding enhancer is to improve the high-order cold-start neighbors' embedding quality.

Here we take group embedding as an example. The embedding enhancer tackles the user/item embedding is similar to the group embedding.
Specifically, the embedding enhancer $f_{meta}$ is instantiated as a self-attention learner~\cite{nips/VaswaniSPUJGKP17}.
For each  group $g$, the embedding enhancer accepts the randomly initialized first-order embeddings $\{\textbf{h}_{i_{GI_1}}^{0}, \cdots, \textbf{h}_{i_{GI_K}}^{0}\}$,  $\{\textbf{h}_{u_{GU_1}}^{0}, \cdots, \textbf{h}_{u_{GU_K}}^{0}\}$ and $\{\textbf{h}_{g_{GG_1}}^{0}, \cdots, \textbf{h}_{g_{GG_K}}^{0}\}$ from the corresponding subgraphs $\mathcal{G}_{GI}$, $\mathcal{G}_{GU}$ and $\mathcal{G}_{GG}$ as input, 
 outputs the smoothed embeddings  $\{\textbf{h}_{i_{GI_1}}, \cdots, \textbf{h}_{i_{GI_K}}\}$,  $\{\textbf{h}_{u_{GU_1}}, \cdots, \textbf{h}_{u_{GU_K}}\}$ and $\{\textbf{h}_{g_{GG_1}}, \cdots, \textbf{h}_{g_{GG_K}}\}$, and use the average function to obtain the meta embedding $\hat{\textbf{h}}_{g_{GI}} $,  $\hat{\textbf{h}}_{g_{GU}} $,  $\hat{\textbf{h}}_{g_{GG}} $.  The process is:

\beqn{
	\label{eq:meta_learner}
	\{ \textbf{h}_{i_{GI_1}}, \cdots, \textbf{h}_{i_{GI_K}} \}  &\leftarrow& \text{SELF\_ATTENTION} ( \{ \textbf{h}_{i_{GI_1}}^{0}, \cdots, \textbf{h}_{i_{GI_K}}^{0} \} ), \nonumber \\ 
	\hat{\textbf{h}}_{g_{GI}} &=& \text{AVERAGE} (\{ \textbf{h}_{i_{GI_1}}, \cdots, \textbf{h}_{i_{GI_K}} \}),  \\ \nonumber 
}

\noindent where the embedding $\hat{\textbf{h}}_{g_{GU}}$ and $\hat{\textbf{h}}_{g_{GG}}$ can be obtained in the same way.
Furthermore, the aggregated group embedding $\textbf{h}_{g_{GU'}}^{L}$ in Eq.~\eqref{eq:fuse}  is also considered to reconstruct the group embedding. 
Finally, $f_{meta}$  fuses these embeddings using Eq.~\eqref{eq:aggregate} to obtain another meta   embedding $\hat{\textbf{h}}_{g}$.

The advantage of the self-attention learner is to pull the similar nodes more closer, while pushing away dissimilar nodes. Thus, the self-attention technique can capture the major group/user/item preference from its neighbors.
Same with~\secref{sec:basic_reconstruction}, the same cosine similarity (i.e., Eq.\eqref{eq:consine_similarity}) is used between $\hat{ \textbf{h}}_g$ and  $\textbf{h}_g$ to measure their similarity. 
Once the embedding enhancer $f_{meta}$ is trained, we incorporate the meta  embedding $\hat{\textbf{h}}_{g_{GI}}$, $\hat{\textbf{h}}_{g_{GU}}$ and $\hat{\textbf{h}}_{g_{GG}}$ produced by the embedding enhancer into the GNN model at each graph convolution step (i.e., we add the meta embedding into Eq.~\eqref{eq:basic_graph_convolution}):
\beqn{
	\label{eq:meta_aggregator}
	\textbf{h}_{g_{GI}}^{l} &=& \text{CONV}( \hat{\textbf{h}}_{g_{GI}}, \textbf{h}_{g_{GI}}^{l-1},   \textbf{h}^{l}_{  \mathcal{N}(g_{GI})}  ), \\ \nonumber
	\textbf{h}_{g_{GU}}^{l} &=&  \text{CONV} (\hat{\textbf{h}}_{g_{GU}}, \textbf{h}_{g_{GU}}^{l-1},  \textbf{h}^{l}_{ \mathcal{N}(g_{GU})} ), \\ \nonumber
	\textbf{h}_{g_{GG}}^{l} &=& \text{CONV}( \hat{\textbf{h}}_{g_{GG}}, \textbf{h}_{g_{GG}}^{l-1},   \textbf{h}^{l}_{  \mathcal{N}(g_{GG})}  ).
}
\quad For a target group $g$, we repeat Eq.~\eqref{eq:meta_aggregator}  $L$ steps to obtain the embeddings $\textbf{h}_{g_{GI}}^{L}$, $\textbf{h}_{g_{GU}}^{L}$,  $\textbf{h}_{g_{GG}}^{L}$. Then, we use Eq.~\eqref{eq:fuse} to obtain the aggregated group embedding $\textbf{h}_{g_{GU'}}^{L}$, and use Eq.~\eqref{eq:aggregate}  to obtain the final group embedding $\textbf{h}_g^L$.  Finally, we also use  cosine similarity (i.e., Eq.~\eqref{eq:consine_similarity}) to optimize the model parameters including  the  GNNs' parameters $ \Theta_{f}$  and the embedding enhancer's parameters $\Theta_{f_{meta}}$. Similarly, the embedding enhancer $f_{meta}$ can obtain the enhanced user embedding on $\mathcal{G}_{UI}$ and $\mathcal{G}_{UU}$, and  the enhanced item embedding on $\mathcal{G}_{UI}$.

%The meta aggregator extends the original GNN graph convolution through emphasizing the representations of the cold-start neighbors in each convolution step, which can improve the final embeddings of the target groups. 

\subsection{Model Training}
\label{sec:multi_task_training}
We conduct multi-task learning paradigm~\cite{conf/sigir/WuWF0CLX21} to optimize the model parameters, i.e., we jointly train the recommendation objective function (cf. Eq.~\eqref{eq:bprloss}) and the designed SSL objective function (cf. Eq.~\eqref{eq:over_all_reconstruction}):
\beqn{
	\label{eq:overall_loss}
	\mathcal{L} = \mathcal{L}_{main} + \lambda_1 \mathcal{L}_{R}  + \lambda_2 ||\Theta||^{2}_{2},
}

\noindent where $\Theta$ = $\{ \Theta_f, \Theta_{f_{meta}} \}$ is the model parameters, $\lambda_1$  and $\lambda_2$ are hyperparameters. We also consider another training paradigm~\cite{hao2021pre}, i.e., pre-train the GNNs on
$\mathcal{L}_{R}$  and fine-tune the GNNs on $\mathcal{L}_{main}$. We detail the recommendation performance of the two training paradigms in~\secref{sec:multitask_vs_pretrain}.

\begin{table*}[t]
	\caption{The time complexity analysis between GNN and \tRC.}
	\label{tab:time-complexity}
		\newcolumntype{?}{!{\vrule width 0.9pt}}
	\newcolumntype{C}{>{\centering\arraybackslash}p{1.0em}}
	
	\centering  \small
	\renewcommand\arraystretch{1.0}
	
	\resizebox{0.75\textwidth}{!}{
		\begin{tabular}{c|c|c}
			\hline
			Component      & GNN & \sRC \\ \hline\hline
						\begin{tabular}[c]{@{}c@{}}Adjacency\\ Matrix\end{tabular}     
			& $ \mathcal{O}(2 | \mathcal{E} |)$                  
			& $\mathcal{O}(10 |\hat{ \mathcal{E} |} s + 2 | \mathcal{E} |)$                \\ \hline
				\begin{tabular}[c]{@{}c@{}}Graph\\ Convolution\end{tabular}    
			& $\mathcal{O}(2 | \mathcal{E} | L d s \frac{| \mathcal{E} |}{B})$                  
			& $\mathcal{O}(2 (| \mathcal{E}| + 5|\hat{\mathcal{E}} |) L d s \frac{| \mathcal{E} |}{B})$                \\ \hline
			\begin{tabular}[c]{@{}c@{}}BPR\\ Objective Function\end{tabular}             
			& $\mathcal{O}(2 | \mathcal{E} | d s)$                  
			& $\mathcal{O}(2 | \mathcal{E} | d s)$                \\ \hline
			\begin{tabular}[c]{@{}c@{}}Self-supervised\\ Objective Function\end{tabular}             
			& -                  
			& $\mathcal{O}(20|\hat{\mathcal{E}}|Lds )$                \\ \hline

	\end{tabular}}
\end{table*}

\subsection{Time and Space Complexity Analysis}
We present the time and space complexity of \RC, and compare its time and space complexity with the backbone GNN model. 
Same as LightGCN~\cite{sigir/0001DWLZ020}, we implement \sRC as the matrix form.
Suppose  the number of edges in the interaction graph $\mathcal{G}$ is  $|\mathcal{E}|$. The number of edges in the masked interaction graph $\hat{\mathcal{G}}$ is  $|\hat{\mathcal{E}}|$. 
Since we mask a large proportion neighbors for each node in $\mathcal{G}$ to simulate the cold-start scenario, the size of  masked edge set is far less than
the original edge set, i.e.,  $|\hat{\mathcal{E}}| \ll |\mathcal{E}|$.
Let $s$ denote the number of epochs, $d$ denote the embedding size and $L$ denote the GCN convolution layers.
Since \sRC introduces meta embedding to enhance the aggregation ability, the space complexity of \sRC is twice than that of the vanilla GNN model. 
The time complexity comes from four parts, namely adjacency matrix normalization, graph convolution operation, recommendation objective function and self-supervised objective function. 
Since we do not change the GNN's model structure and GNN's inference process, the time complexity of \sRC in the graph convolution operation and recommendation objective function is the same as the vanilla GNN model.  We present the main differences between the vanilla GNN and \sRC models as follows:

\begin{itemize}[ leftmargin=10pt ]
	\item Adjacency matrix normalization. In each training epoch, generating the target group embedding needs five corresponding subgraphs. Suppose the non-zero elements in the adjacency matrices of the full training graph and the five subgraphs are $2|\mathcal{E}|$, $2| \hat{\mathcal{E}}_{UU}|$, $2| \hat{\mathcal{E}}_{UI}|$, $2| \hat{\mathcal{E}}_{GG}|$, $2| \hat{\mathcal{E}}_{GU}|$ and $2| \hat{\mathcal{E}}_{GI}|$, respectively. Thus, the total complexity of adjacency matrix normalization is 
	$\mathcal{O}( (2|\hat{\mathcal{E}}_{UU}| + 2|\hat{\mathcal{E}}_{UI}| + 2|\hat{\mathcal{E}}_{GG}| + 2|\hat{\mathcal{E}}_{GU} | + 2|\hat{\mathcal{E}}_{GI}|)s  + 2 | \mathcal{E}|) \approx 10 |\hat{\mathcal{E}}|s + 2 |\mathcal{E}|$.~\footnote{Notation $|\hat{\mathcal{E}}| $ denotes the masked edge length $| \hat{\mathcal{E}}_{UU}|$, $| \hat{\mathcal{E}}_{UI}|$, $| \hat{\mathcal{E}}_{GG}|$, $| \hat{\mathcal{E}}_{GU}|$ and $| \hat{\mathcal{E}}_{GI}|$.}
	
	\item Self-supervised objective function. We evaluate the self-supervised tasks upon the masked subgraphs. For the user or item embedding reconstruction task, the time complexity is $\mathcal{O}(2d *( 2|\hat{\mathcal{E}}_{UU}| + 2|\hat{\mathcal{E}}_{UI}| )* s * L) \approx 8|\hat{\mathcal{E}}|Lds $. For the group embedding reconstruction task, the time complexity is $\mathcal{O}(2d *(  2|\hat{\mathcal{E}}_{GG}| + 2|\hat{\mathcal{E}}_{GU}| + 2|\hat{\mathcal{E}}_{GI}| )* s * L) \approx 12|\hat{\mathcal{E}}|Lds $, where $2d$ represents the concatenated embedding size, as we incorporate the meta embedding to the graph convolution process. Thus the total time complexity of self-supervised loss is $ 8|\hat{\mathcal{E}}|Lds + 12|\hat{\mathcal{E}}|Lds =    20|\hat{\mathcal{E}}|Lds $.

\end{itemize}
We summarize the time complexity between the vanilla GNNs and \sRC in Table 1, from which we observe that the time complexity of \sRC is in the same magnitude with the vanilla GNNs, which is totally acceptable, since the increased time complexity
of \sRC is only from the self-supervised loss. The details are shown in~\secref{sec:effect_meta_learning}.
%In practice, taking the Weeplaces data as an example, the time complexity of \sRC is about 2.7x larger than the vanilla GNN model, which is totally acceptable considering the speedup of convergence speed we will show in~\secref{sec:effect_meta_learning}.
%The testing platform is Tesla P40 graphics card equipped with Inter i5-2640 CPU (32GB Memory). The time cost of each epoch on Weeplaces is 179.2s and 483.84s for LightGCN (alternative) and \sRC respectively, which is consistent with the complexity analyses.

\hide{
\begin{algorithm}[t]
	{\small \caption{The Overall Training Process. \label{algo:rl}}
		Pre-train the meta aggregator with parameter $\Theta_{m_1}$; \\
		Pre-train the GNN model with parameter $\Theta_f$, $\Theta_{m_1}$;\\
		Maximize the group-user mutual information with parameter $\Theta_{m_2}$; \\
		Fine tune the trained parameter $\Theta_f$, $\Theta_{m_1}$ and $\Theta_{m_2}$ into the downstream group recommendation task.
		
}\end{algorithm}
}

%% file: exp.tex
\section{Experiments}
\label{sec:experiment}
We conduct comprehensive experiments to answer the following questions:

\begin{itemize}[ leftmargin=10pt ]
	\item \textbf{Q1:} Can \sRC have better performance than the other baselines?
		
	\item \textbf{Q2:} Can the proposed self-supervised task benefit the occasion group recommendation task?
	
	\item \textbf{Q3:} How does \sRC perform in different settings?
\end{itemize}

\subsection{Experimental Settings}

\subsubsection{Datasets.}
We select  three public recommendation datasets, i.e., Weeplaces~\cite{sigir/SankarWWZYS20}, 
CAMRa2011~\cite{sigir/Cao0MAYH18} and Douban~\cite{YinW0LYZ19} to evaluate the performance of \RC. Table 1 shows
the statistics of these three datasets. 

\begin{table*}[t]
	\newcolumntype{?}{!{\vrule width 0.9pt}}
	\newcolumntype{C}{>{\centering\arraybackslash}p{1.0em}}
	\caption{
		\label{tb:statistics} Statistics of the Datasets. 
		\normalsize
	}
	
	\centering  \small
	\renewcommand\arraystretch{1.0}
	
	\begin{tabular}{crrr } 
		\toprule
		&  Weeplaces & CAMRa2011  & Douban 
		\\
		\midrule
		Users
		&8,643 & 602 &  70,743
		\\
	 Items
		&25,081  & 7,710 & 60,028
		\\
	   Groups
		& 22,733 & 290 & 109,538
		\\
		 U-I Interactions
	 	& 1,358,458	 & 116,314  &  3,422,266
		\\
		 G-I Interactions
		&  180,229& 145,068 &  164,153
		\\
		U-I Sparsity
		& 6.27\% & 2.51\%  & 0.081\%  
		\\
		G-I Sparsity
		& 0.03\% & 6.49\%  &  0.002\%
		\\
		\bottomrule
	\end{tabular}
	
\end{table*}

\subsubsection{Baselines.}
We select the following baselines:

\begin{itemize}[ leftmargin=10pt ]
	\item \textbf{MoSAN~\cite{sigir/TranPTLCL19}} adopts sub-attention mechanism to model the group-item interactions.
	
	\item \textbf{AGREE~\cite{sigir/Cao0MAYH18}} adopts attention mechanism for jointly modelling user-item and group-item interactions.
		
	\item \textbf{SIGR~\cite{YinW0LYZ19}} further incorporates social relationships of groups and users to model the attentive group and user representations.
	
	\item \textbf{GroupIM~\cite{sigir/SankarWWZYS20}} further regularizes  group and user representations by maximizing the mutual information between the group and its members.
	
	\item \textbf{GAME~\cite{sigir/HeCZ20}} performs graph convolution only based on the first-order neighbors from the group-group, group-user and group-item graphs for group recommendation.
	
	\item \textbf{GCMC~\cite{BRDTKDD2018}} uses the classical GCN~\cite{Thomasgcn} model to perform graph convolution, and learn the node embeddings.
	
	\item \textbf{NGCF~\cite{wangncgf19}} adds second-order interactions upon the neural passing based GNN model~\cite{GSSPPMLR17}. 
	
	\item \textbf{LightGCN~\cite{sigir/0001DWLZ020}} devises the light graph convolution upon NGCF. 
	
	\item \textbf{HHGR~\cite{conf/CIKM21/JMJLJH21}} designs coarse- and fine-grained node dropout strategies upon the hypergraph for group recommendation. 
\end{itemize}

We discard potential baselines like Popularity~\cite{conf/recsys/CremonesiKT10}, COM~\cite{kdd/YuanCL14}, and CrowdRec~\cite{conf/wsdm/RakeshLR16}, since previous works~\cite{sigir/TranPTLCL19,sigir/Cao0MAYH18,YinW0LYZ19,sigir/SankarWWZYS20,sigir/HeCZ20} have validated the superiority over the compared ones. 
For the GNN model GCMC, NGCF and LightGCN, we  extend it to address group recommendation as proposed in~\secref{sec:base_gnn}; Besides, notation GNN* denotes the corresponding proposed  model \RC. We further evaluate two variants of  \RC, named Basic-GNN and Meta-GNN, which are  equipped with  basic embedding reconstruction with GNNs (\secref{sec:basic_reconstruction}) and embedding enhancer (\secref{sec:meta_reconstruction}), respectively.

\subsubsection{Training Settings.} 
\label{sec:training_settings}

We present the details of dataset segmentation, model training process and hyper-parameter settings.

\vpara{Dataset Segmentation.}
We first select the groups with abundant interactions as the target groups in the meta-training set $D_T^g$ , and leave the rest groups in the meta-test set $D_N^g$, as we need more accurate embeddings of groups to evaluate the  quality of the generated group embeddings.
The splitting strategy of the users/items is the same with the groups.
In order to avoid information leakage, we further select items with sufficient interactions from the group meta-training set $D_T^g$ and the user meta-training set $D_T^u$, and obtain the meta-training set $D_T^i$.
For simplicity, we use $D_T$ and $D_N$ to denote these meta-training and meta-test sets.
For each group/user in $D_N$, according to the interaction time with items, we further choose the former  $c$\%  items  into the training set  $Train_N$, and leave
the other items into the test set $Test_N$.

More concretely, for addressing the  occasional groups, we split the dataset according to a predefined hyperparameter $n_g$. If a group interacts with more than $n_g$ items, we put it in $D_T$. Otherwise, we leave the group in $D_N$. We select $n_g$ as 10  for Weeplaces and Douban.
Similarly, for the users/items with few interactions, we split the dataset according to $n_u$ ($n_i$). Both $n_u$ and $n_i$ is set as 10 for Weeplaces and Douban.
In CAMRa2011, since the groups, users and items have abundant interactions, we randomly select 70\% groups, users and items in $D_T$ and leave the rest in $D_N$.
For each group and user in $D_N$, in order to simulate the real cold-start scenario, we only keep top 10  interacted items in chronological order.
Similarly, for each item in $D_N$, we only keep its first 5 interacted groups/users.
 
\hide{Since \sRC explicitly considers the user-user and group-group connections, for these three datasets, we build the user-user subgraph $\mathcal{G}_{UU}$ and group-group subgraph $\mathcal{G}_{GG}$ according to the number of shared items. In $\mathcal{G}_{UU}$, if two users share with more than 30 items, we add a edge between them; In $\mathcal{G}_{GG}$, if two groups share with more than 10 items, we add a edge between them.}

\vpara{Model Training Process.}
We train each of the baseline methods to obtain the ground-truth embeddings on $D_T$, since these methods can learn high-quality embeddings for the target nodes with enough interactions.
%In order to verify this, we  report the recommendation performance using ground-truth embeddings  generated by different baseline methods, and explore whether \sRC is sensitive to these embeddings in~\secref{sec:sensitive_ground_truth}.
For MoSAN, AGREE, SIGR, GroupIM and GAME, we directly fetch the trained embeddings as ground-truth embeddings;  For the hypergraph GNN model HHGR and the general GNNs (i.e., LightGCN, NGCF and GCMC), we first fetch the embeddings at each layer and then concatenate them to obtain the final ground-truth embeddings.
Take the group embedding as an example, $\textbf{h}_g$ = $ \textbf{h}_g^{0}$   + $\cdots$+  $\textbf{h}_g^L$. User and item embeddings can be explained in the similar way.
%For each group/user in $D_N$,we further select top $c$ of its/his interacted items in chronological order into the training set  $Train_N$, and leave the rest items into the test set $Test_N$.
%Although \sRC is trained under the multi-task learning paradigm, the dataset we used to train the SSL tasks and the main recommendation task is a little bit different.

The SSL tasks is trained on $D_T$, while the recommendation task is trained on $D_T$ and $Train_N$. Both the SSL and the recommendation tasks are evaluated in $Test_N$.
We adopt Recall@$\mathcal{K}$ and NDCG@$\mathcal{K}$ as evaluation metric.
\hide{Note that in order to avoid the information leakage, we select the items with sufficient interactions only in the group-item and user-item training files.}

\vpara{Hyper-parameter Settings.}
We use Xavier method~\cite{jmlr/GlorotB10} to initialize the parameters of all the models.
We set the learning rate as 0.001 and the mini-batch size as 256.
We tune $K$, $L$, $c\%$ within
the ranges of \{3, 4, 5, 6, 7, 8, 9, 10, 11, 12\}, \{1,2,3,4\} and \{0.1, 0.2, 0.3\}, respectively. We tune $\lambda_1$ with the ranges of \{0.01, 0.1, 0.5, 1.0, 1.2\}, and empirically set $\lambda$ and $\lambda_2$ as 1 and 1e-6, respectively. We tune $c_u$ and $c_g$ with the ranges of \{10, 20, 30\}.
By default, we set $L$ as 3, $K$ as 5, $c$\% as 0.1, $\tau$ as 0.2, $c_u$ and $c_g$ as 20, and $\mathcal{K}$ as 20.
%For the unique ones of \RC,
%The models are optimized by the Adam optimizer
%The early stopping strategy is the same as LightGCN.
%The proposed \sRC inherit the optimal values of the shared hyperparamters.
%We tune the self-attention blocks and the self-attention heads in the meta aggregator as \{1, 2, 3, 4\} and \{2, 4, 6, 8\}, respectively; 

\begin{table}[t]
	\newcolumntype{?}{!{\vrule width 1pt}}
	\newcolumntype{C}{>{\centering\arraybackslash}p{3.9em}}
	\caption{
		\label{tb:recommendation} Overall  performance with  sparse rate $c\%$=0.1, layer size $L$=3 and neighbor size $K$=5.
		\normalsize
	}
	\centering  \scriptsize
	\renewcommand\arraystretch{1.0}
	\begin{tabular}{@{~}l@{~}?*{1}{CC?}*{1}{CC?}*{1}{CC}}
		\toprule
		\multirow{2}{*}{\vspace{-0.3cm} Methods }
		&\multicolumn{2}{c?}{Weeplaces}
		&\multicolumn{2}{c?}{CAMRa2011}
		&\multicolumn{2}{c}{Douban}
		\\
		\cmidrule{2-3} \cmidrule{4-5} \cmidrule{6-7}
		& {Recall} & {NDCG} & {Recall} & {NDCG} &{Recall} &{NDCG} \\
		\midrule 
		${\rm MoSAN}$
		& 0.0223& 0.0208
		&0.0214 & 0.0166 
		&0.0023 &0.0019 
		\\
		${\rm AGREE}$
		& 0.0266& 0.0233
		&0.0237 & 0.0168
		&0.0024 & 0.0018 
		\\
		${\rm SIGR}$
		&0.0276&0.0223
		&0.0278 & 0.0169 
		&0.0028 & 0.0021
		\\
		${\rm GroupIM}$
		&0.0228&0.0283
		&0.0277 &0.0169 
		&0.0034 & 0.0026 
		\\
		\midrule
		${\rm GAME}$
		&0.0283&0.0216
		&0.0499 & 0.0173 
		&0.0031 & 0.0027
		\\
		${\rm GCMC}$
		& 0.0312&	0.0083
		&0.0348 &0.0171
		&0.0036 &0.0025
		\\
		${\rm NGCF}$
		&0.0336 & 0.0093
		&0.0288 & 0.0177
		&0.0043&0.0028
		\\
		${\rm LightGCN}$
		&0.0316&0.0233
		&0.1036 & 0.0183
		& 0.0118& 0.0032
		\\
		${\rm HHGR}$
		&0.0488&0.0422
		&0.1494&0.0376
		&0.0154&0.0045
		\\
		\midrule
		${\rm GCMC*}$
		&\textbf{0.0513}&\textbf{0.0448}
		&\textbf{0.1112}&\textbf{0.0394}
		&\textbf{0.0053}&\textbf{0.0038}
		\\
		${\rm NGCF*}$
		& \textbf{0.0486}&\textbf{0.0413}
		&\textbf{0.1256} &\textbf{0.0342}
		&\textbf{0.0133}& \textbf{0.0043}
		\\
		${\rm LightGCN*}$
		&\textbf{0.0523}& \textbf{0.0426}
		&\textbf{0.1634}&\textbf{0.0353}
		&\textbf{0.0237}&\textbf{0.0063}
		\\
		\bottomrule
	\end{tabular}
	
\end{table}

\subsection{Recommendation Performance (Q1)}
\subsubsection{Overall  Recommendation Performance}

We report the overall group recommendation performance in Table~\ref{tb:recommendation}. The results show  that  \sRC (denoted as ${\rm GNN}^{*}$)  has the best recommendation performance, which indicates the proposed SSL tasks are useful to learn high-quality embeddings, and can further benefit the  recommendation task. Besides, \sRC is better than the  most competitive baseline method HHGR, which indicates the superiority of the proposed SSL tasks in dealing with high-order cold-start neighbors. 

\hide{
\begin{itemize}[ leftmargin=10pt ]
	
	%\item Among all the attentive models (MoSAN, AGREE, SIGR, and GroupIM), MoSAN and AGREE perform the worst, as these two methods do not explicitly model the group-user, group-group and user-user connections.

	%\item The vanilla GNN models (GAME, GCMC, NGCF and LightGCN) achieve better performance compared with the attentive models, which indicates incorporating high-order collaborative signals is useful to improve the recommendation performance. Besides, GAME performs the worst among all the vanilla GNN models, which indicates only incorporating the first-order neighbors is not enough to learn high-quality embeddings.

	\item 	Compared with the attentive models (MoSAN, AGREE, SIGR, and GroupIM) and the vanilla GNN models (GAME, GCMC, NGCF and LightGCN), our proposed \sRC (denoted as ${\rm GNN}^{*}$)  significantly improves the
	%recommendation performance, which indicates the proposed SSL tasks are useful to learn high-quality embeddings, and can further benefit the  recommendation task.
	% (+1.19\%-5.98\% in terms of	recall@20)

\end{itemize}
}

\subsubsection{Interacted Number and Sparse Rate Analysis.}
\label{sec:i_s_analysis}
\begin{figure}
	\centering
	
	\mbox{ 
		
		\subfigure[\scriptsize  Weeplaces
		]{\label{subfig:w_num}
			\includegraphics[width=0.45 \textwidth]{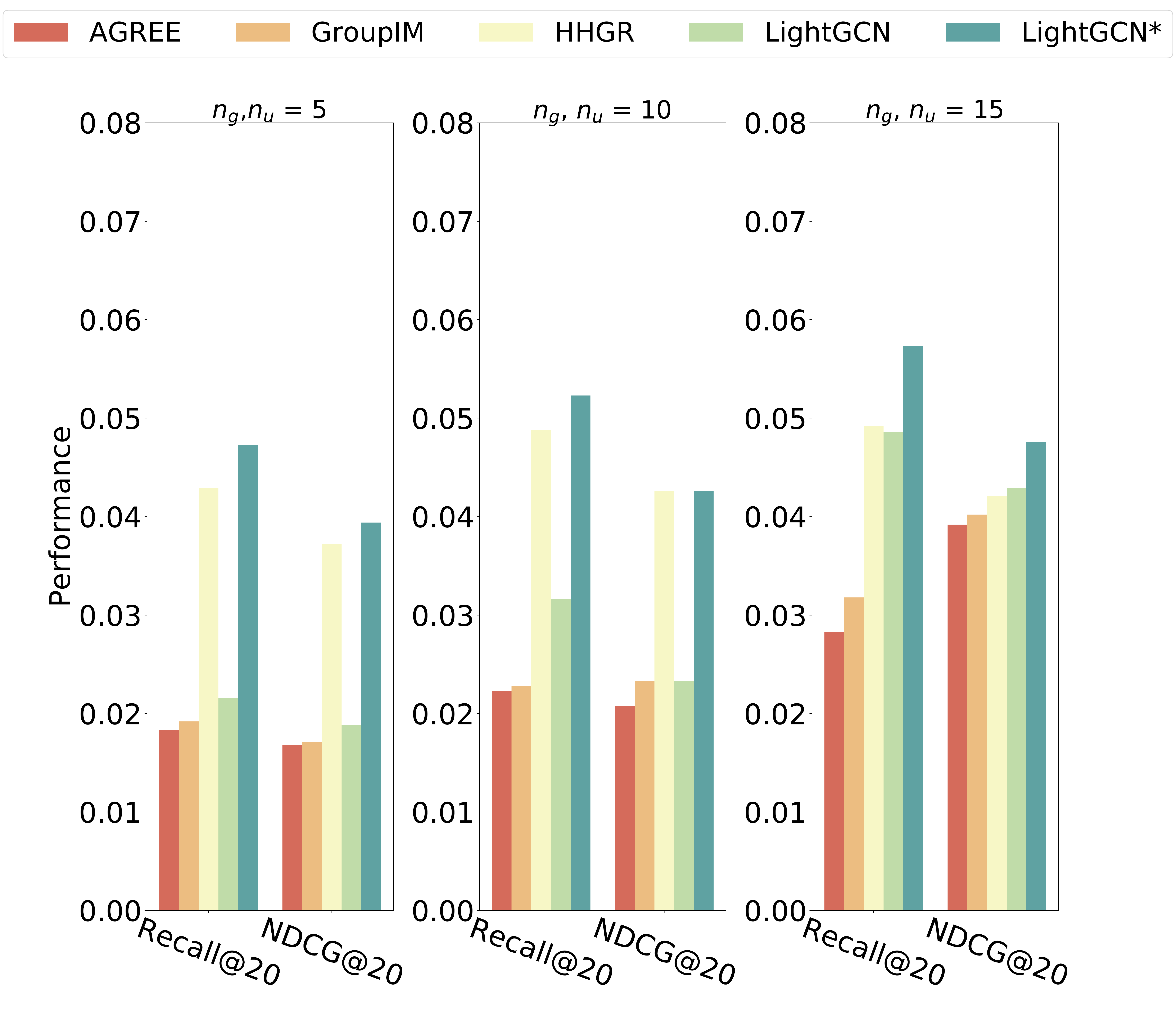}
		}
		
		\hspace{-0.2cm}
		
		\subfigure[\scriptsize  Weeplaces
	]{\label{subfig:w_r}
		\includegraphics[width=0.45 \textwidth]{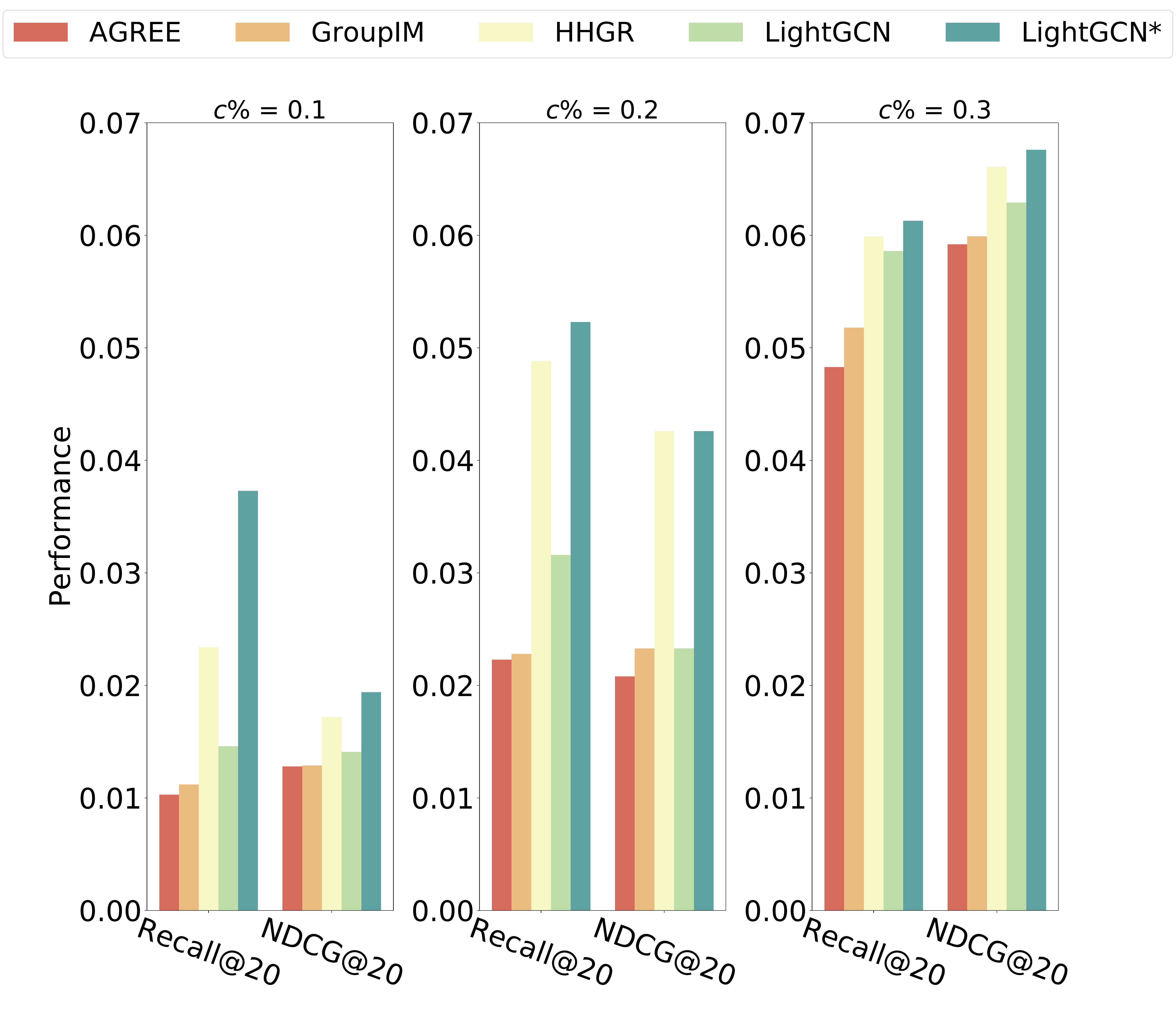}
	}
		
	}

	\mbox{ 
	
	\subfigure[\scriptsize  CAMRa2011
	]{\label{subfig:a_c_num}
		\includegraphics[width=0.45\textwidth]{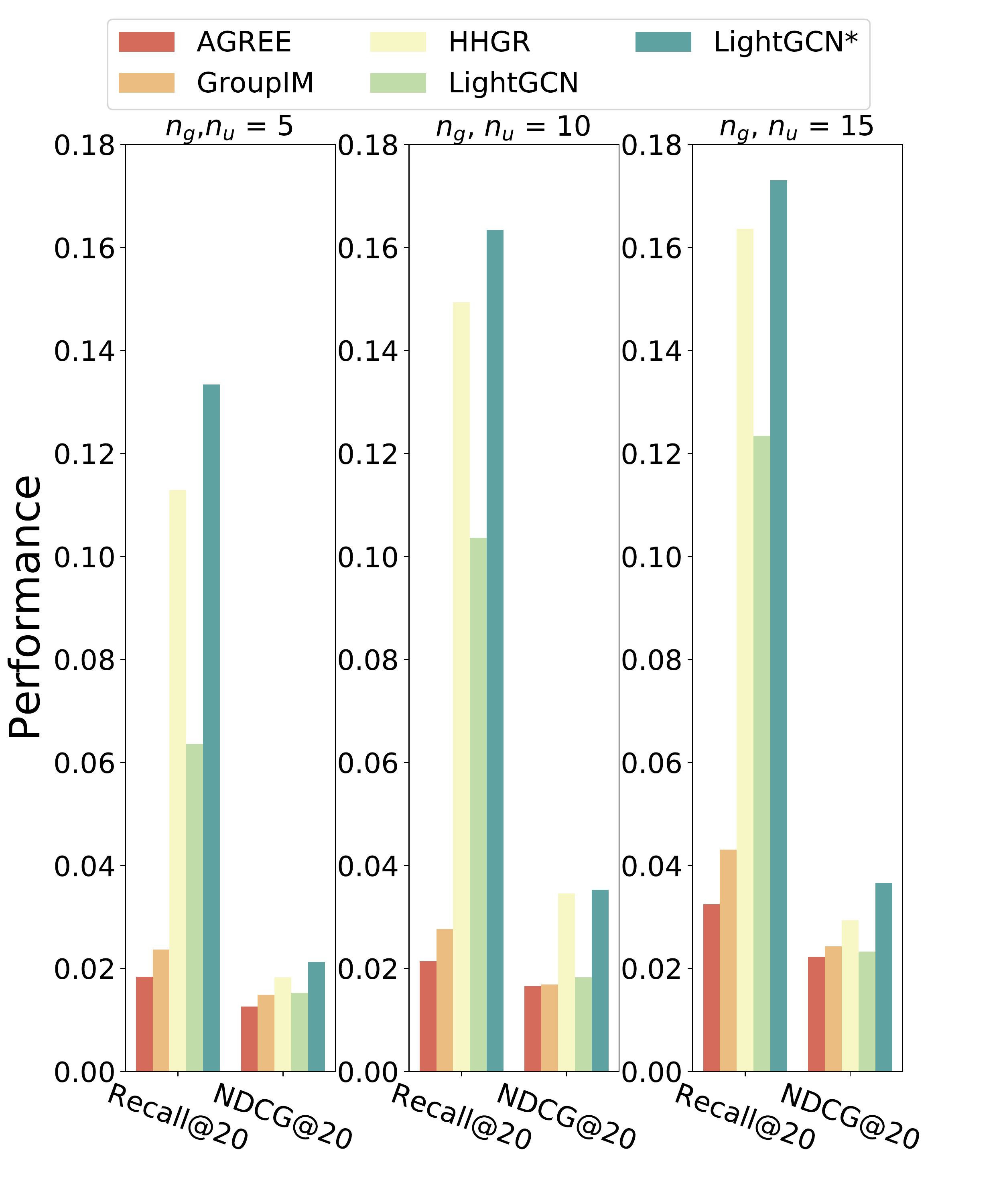}
	}

	\subfigure[\scriptsize  CAMRa2011
	]{\label{subfig:a_c_r}
		\includegraphics[width=0.45\textwidth]{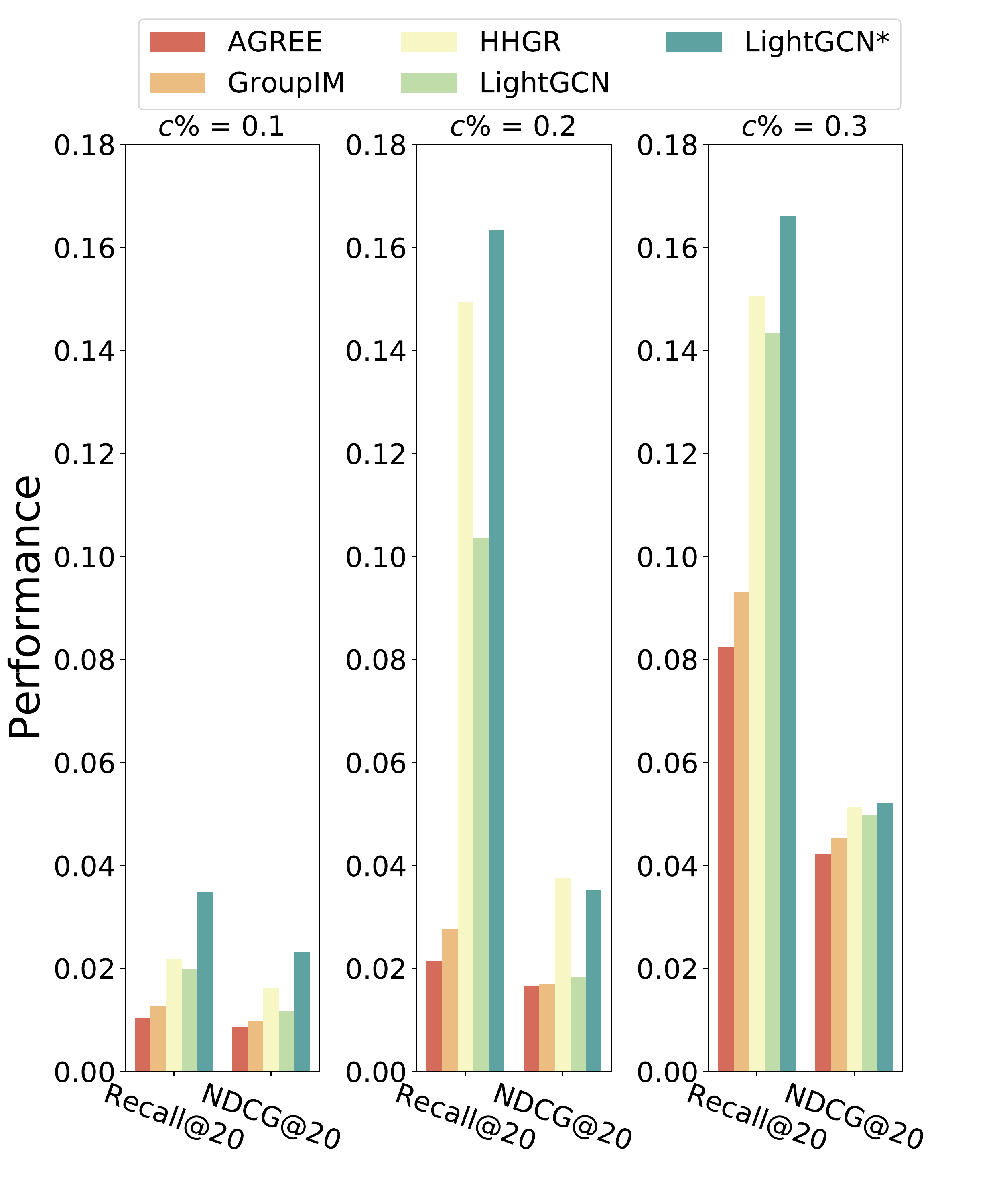}
	}
	
}

\mbox{ 
	
	\subfigure[\scriptsize  Douban
	]{\label{subfig:a_d_num}
		\includegraphics[width=0.45\textwidth]{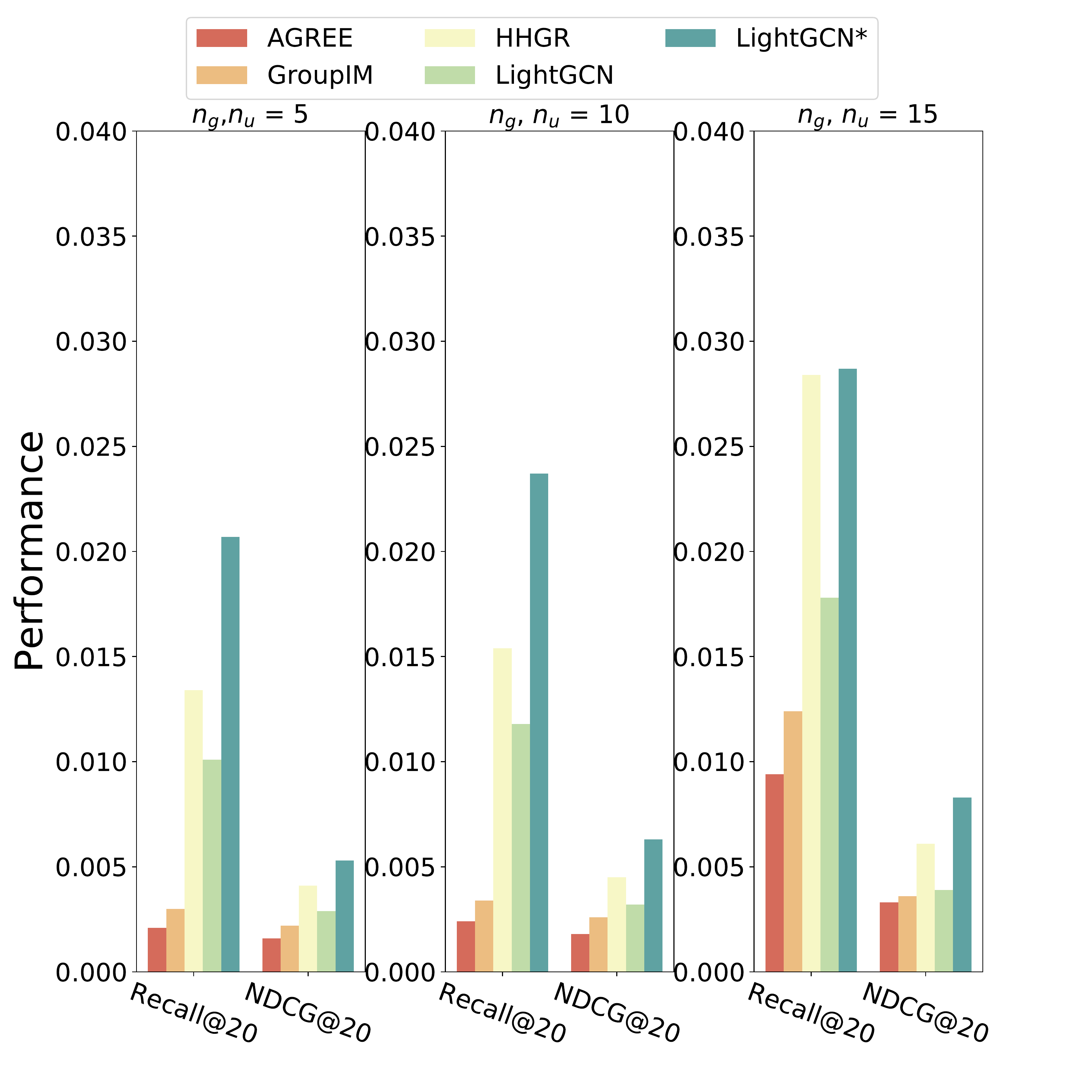}
	}

	\subfigure[\scriptsize  Douban
	]{\label{subfig:a_d_r}
		\includegraphics[width=0.45\textwidth]{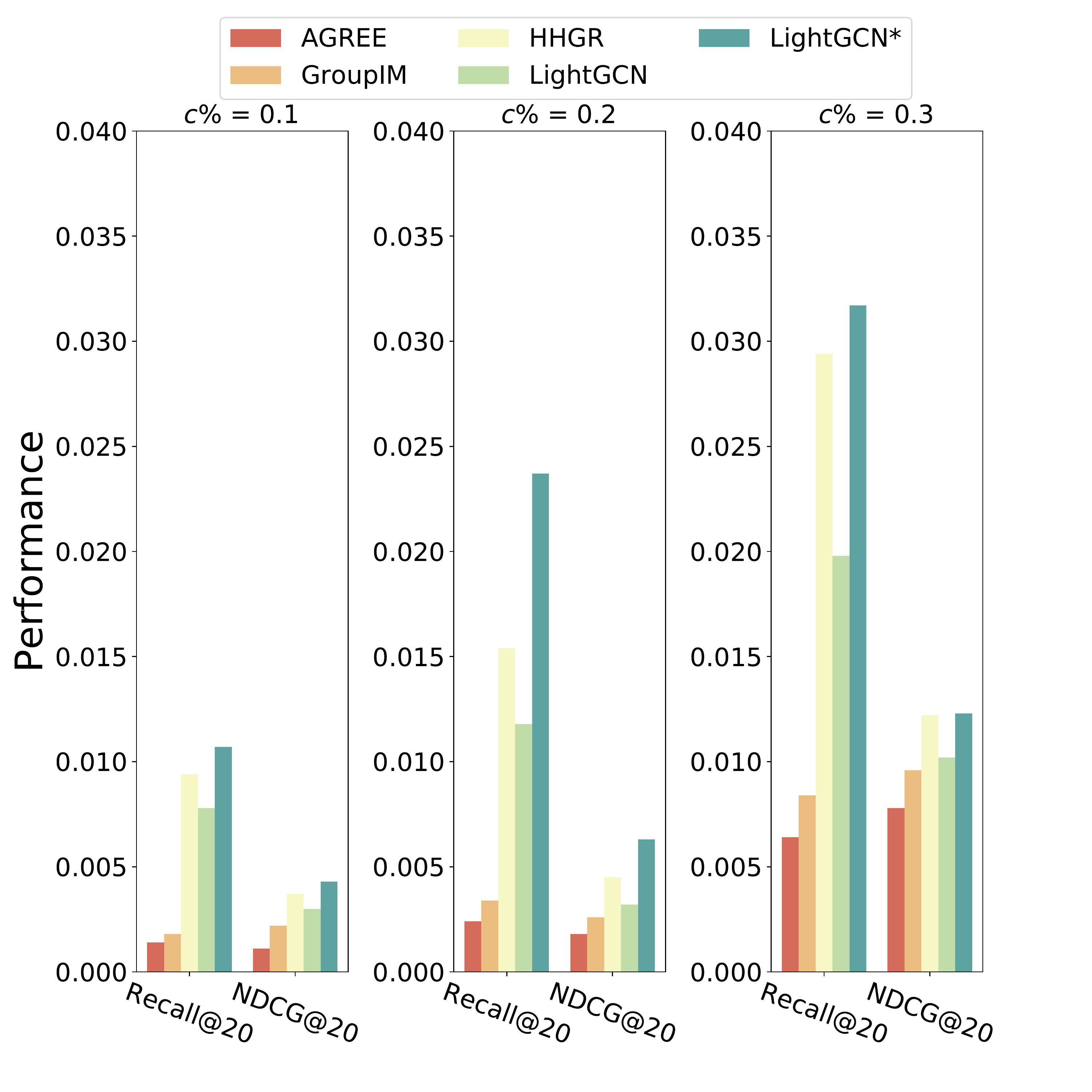}
	}
	
}

	\caption{\label{fig:num} \small{Recommendation performance under different interacted numbers $n_u$ and $n_g$ (Cf. Fig~\ref{subfig:w_num}, Fig~\ref{subfig:a_c_num} and Fig~\ref{subfig:a_d_num}), and under different sparse rate $c$\% (Cf. Fig~\ref{subfig:w_r}, Fig~\ref{subfig:a_c_r} and Fig~\ref{subfig:a_d_r}).}}
\end{figure}

Since we split the groups/users into $D_T$ and $D_N$ according to the 
predefined hyperparameters  $n_g$ and $n_u$ or the sparse rate $c$\%, in order to explore whether \sRC is sensitive to these hyperparameters, 
we change $n_g$ and $n_u$ in the range of $\{5, 10, 15  \}$ while keeping $c$\% as 0.1, $L$ as 3 and $K$ as 5; and change 
$c\%$ in the range of $\{ 0.1, 0.2, 0.3 \}$ while keeping $n_g$ and $n_u$ as 5, $L$ as 3 and $K$ as 5.  
We compare our proposed model \sRC (denoted as LightGCN*, in which we select LightGCN as the backbone GNN model) with competitive baseline methods AGREE, GroupIM, HHGR and LightGCN, and report the  recommendation performance in Figure~\ref{fig:num}. If $n_g$, $n_u$ and $c$\% gets smaller, the groups and users in $D_N$ will have fewer interacted items. 
The results show that:  (1) ${ \rm LightGCN}^{*}$ has the best performance, which shows \sRC is able to handle the cold-start recommendation with different $n_g$, $n_u$ and $c$\%. 
 (2) When $n_g$ and $n_u$ decrease from 15 to 5, and  when $c$\% decreases from 0.3 to 0.1, compared with other baselines,  \sRC always has a large improvement, which also shows its capability in dealing with the cold-start group recommendation problem.

\subsection{Ablation Study (Q2)}
We perform the ablation study to explore whether each component in \sRC is reasonable for the good recommendation performance.
To this end, we report the recommendation performance for \sRC and its variant model  in Table~\ref{tb:ablation}.
We find that: (1) Basic-GNN and Meta-GNN is consistently superior than the vanilla GNNs, which indicates the effectiveness of the proposed SSL tasks. 
(2) Among all the variant models, Meta-GNN performs the best, which  indicates  enhancing the cold-start neighbors' embedding quality is much more important. (3) GNN* performs the  best, which verifies the superiority of combining these  SSL tasks.

\hide{
\begin{itemize}[ leftmargin=10pt ]

\item Basic-GNN, Meta-GNN and CL-GNN is consistently superior than the vanilla GNNs, which indicates the effectiveness of the proposed SSL tasks. 

\item Among Basic-GNN, Meta-GNN, CL-GNN, Meta-GNN performs the best, which  indicates  improving the embedding quality of  cold-start neighbors is much more important.

%CL-GNN performs better than the vanilla GNN models by improving 0.02-0.27\% in terms of Recall@20, which shows the effectiveness of explicitly considering the correlations between the group and its members.
	
%\item  All the Basic-GNN models beat the corresponding vanilla GNN models by improving  0.09-2.13\% in terms of Recall@20, which indicates  the  
%basic embedding reconstruction task is capable of learning high-quality embeddings.

%\item Compared with the Basic-GNN models, the Meta-GNN models can further improve the recommendation performance by 0.08-2.01\% in terms of Recall@20,  which indicates  the meta aggregator can indeed strengthen each layer's aggregation ability.

%\item Compared with Basic-GNN and Meta-GNN, CL-GNN performs the worst, which implies improving the quality of cold-start embedding is much more critical than only considering the associations between the group and its members.

\item GNN*  leads the best performance, which verifies the superiority of the proposed  SSL tasks.

\end{itemize}
}

\begin{table}[t]
	\newcolumntype{?}{!{\vrule width 1pt}}
	\newcolumntype{C}{>{\centering\arraybackslash}p{3.4em}}
	\caption{
		\label{tb:ablation} Ablation study for \stRC with  sparse rate $c$\%= 0.1, layer size $L$=3 and neighbor size $K$=5. 
		\normalsize
	}
	\centering  \scriptsize
	\renewcommand\arraystretch{1.0}
	\begin{tabular}{@{~}l@{~}?*{1}{CC?}*{1}{CC?}*{1}{CC}}
		\toprule
		\multirow{2}{*}{\vspace{-0.3cm} Methods }
		&\multicolumn{2}{c?}{Weeplaces}
		&\multicolumn{2}{c?}{CAMRa2011}
		&\multicolumn{2}{c}{Douban}
		\\
		\cmidrule{2-3} \cmidrule{4-5} \cmidrule{6-7}
		& {Recall} & {NDCG} & {Recall} & {NDCG} &{Recall} &{NDCG} \\
		\midrule
		${\rm GCMC}$
		& 0.0312&	0.0083
		&0.0348 &0.0171
		&0.0036 &0.0025
		\\
		${\rm Basic}$-${\rm GCMC}$
		& 0.0412&0.0293
		&0.0561 &0.0278
		& 0.0045& 0.0028
		\\
		${\rm Meta}$-${\rm GCMC}$
		&0.0441&0.0328
		& 0.0826&0.0319
		&0.0048&0.0031
		\\
		${\rm GCMC*}$
		&\textbf{0.0513}&\textbf{0.0448}
		&\textbf{0.1112}&\textbf{0.0394}
		&\textbf{0.0053}&\textbf{0.0038}
		\\
		\midrule
		${\rm NGCF}$
		&0.0336 & 0.0093
		&0.0288 & 0.0177
		&0.0043&0.0028
		\\
		${\rm Basic}$-${\rm NGCF}$
		& 0.0390&0.0241
		&0.0971&0.0233
		&0.0068 & 0.0041
		\\
		${\rm Meta}$-${\rm NGCF}$
		&0.0480&0.0382
		&0.1172 &0.0319
		&0.0121&0.0042
		\\
		${\rm NGCF*}$
		& \textbf{0.0486}&\textbf{0.0413}
		&\textbf{0.1256} &\textbf{0.0342}
		&\textbf{0.0133}& \textbf{0.0043}
		\\
		\midrule
		${\rm LightGCN}$
		&0.0316&0.0233
		&0.1036 & 0.0183
		& 0.0118& 0.0032
		\\
		${\rm Basic}$-${\rm LightGCN}$
		& 0.0373 & 0.0318
		&0.1252 & 0.0210
		& 0.0181& 0.0038
		\\
		${\rm Meta}$-${\rm LightGCN}$
		&0.0475 & 0.0415
		&0.1556 & 0.0312
		&0.0232&0.0049
		\\
		${\rm LightGCN*}$
		&\textbf{0.0523}& \textbf{0.0426}
		&\textbf{0.1634}&\textbf{0.0353}
		&\textbf{0.0237}&\textbf{0.0063}
		\\
		\bottomrule
	\end{tabular}
\end{table}

\subsection{Study of \stRC (Q3)}

\subsubsection{Effectiveness of Meta-Learning Setting.}
\label{sec:effect_meta_learning}
As mentioned in~\secref{sec:basic_reconstruction}, we train \sRC under the meta-learning setting.
In order to examine whether the meta-learning setting can benefit the recommendation performance and  have satisfactory time complexity,
we compare \sRC and the  vanilla GNN model with a variant model \RC-M, which removes the meta-learning setting. More Concretely, in \RC-M, for each group/user/item, we do not sample $K$ neighbors, but instead directly using their first-order and high-order neighbors to perform graph convolution. We 
report the average recommendation performance, the average training time in each epoch and the average convergent epoch in Table~\ref{tb:impact_meta_learning}. Based on the results, we find that \sRC is consistently superior than \RC-M, and has much more smaller training time in each epoch, much faster converges speed. This indicates training \sRC in the meta-learning setting can not only improve  the model performance, but also improve the  training efficiency.

\hide{
\begin{itemize}[ leftmargin=10pt ]
	\item  \sRC is consistently superior than \RC-M, which indicates training \sRC equipped with meta-learning setting can make the model easily adapted to new occasional groups.
	
	\item   \sRC is consistently superior than \RC-M, and has much more smaller training time in each epoch, much faster converges speed. This indicates training \sRC in the meta-learning setting can not only improve  the model performance, but also improve the  training efficiency and make the model easily and rapidly being adapted to new occasional groups.
	
	%can accelerate the training speed of the model.  Although the vanilla GNNs have the fastest training speed in each epoch, they perform the worst as they can not handle the cold-start embeddings.
	
	%\item  Compared with \RC-M and the vanilla GNNs, \sRC  converges earlier than them, which  shows  training \sRC in the meta-learning setting can also improve the training efficiency, making the model easily and rapidly being adapted to new occasional groups.

\end{itemize}
}

\begin{table*}[t]
	\caption{Recommendation performance, training time per epoch and convergent epochs  w/wo meta-learning setting. } 
	\resizebox{0.90\textwidth}{!}{
		\begin{tabular}{l|cccc|cccc}
			\hline
			\multicolumn{1}{c|}{Dataset} & \multicolumn{4}{c|}{Weeplace} & \multicolumn{4}{c}{CAMRa2011} \\ \hline
			\multicolumn{1}{c|}{Method}  & Recall    & NDCG   & Time  & Epoch & Recall     & NDCG    &Time & Epoch  \\ \hline\hline
			GCMC       & 0.0312           &  0.0083     &  \textbf{188.6s}    & 31    & 0.0348 & 0.0171 & \textbf{51.8s} &       30            \\
			GCMC*-M                           &  0.0509          & 0.0453      & 721.3s   & 30    &   0.1021        & 0.0382      &  172.6s & 22    \\
			GCMC*             & \textbf{0.0513}      & 0.0448   &  499.6s  &  \textbf{12} & \textbf{0.1112}    & \textbf{0.0394}   & 112.3s     & \textbf{10}    \\ \hline
			NGCF           & 0.0336    &  0.0093        & \textbf{182.6s}  & 30   & 0.0288  & 0.0177   & \textbf{58.7s} & 26        \\
			NGCF*-M         & 0.0465          & 0.0403         & 700.2s  & 20  & 0.1123             & 0.0325    & 166.3s        &  13 \\
			NGCF*            & \textbf{0.0486}    &\textbf{0.0413}   & 489.7s  & \textbf{8}  &\textbf{0.1256}     & \textbf{0.0342} & 100.8s & \textbf{8}  \\ \hline
			LightGCN      & 0.316          &  0.0233       & \textbf{179.2s}  &  30  & 0.1036  & 0.0183  & \textbf{48.3s} & 30        \\
			LightGCN*-M       & 0.0511           & 0.0402      & 683.6s   & 18   &   0.1435           &   0.0329      & 153.8s  & 18  \\
			LightGCN*                                &\textbf{0.0523}     &\textbf{0.0426}  &483.4s  & \textbf{10} &\textbf{0.1634}     &\textbf{0.0353}  & 99.3s & \textbf{6}    \\ \hline
	\end{tabular}}
	\label{tb:impact_meta_learning}
\end{table*}

\begin{figure}[t]
	\centering

	\mbox{ 
	
	\subfigure[\scriptsize  Recommendation performance
	]{\label{subfig:gt_lightgcn}
		\includegraphics[width=0.45 \textwidth]{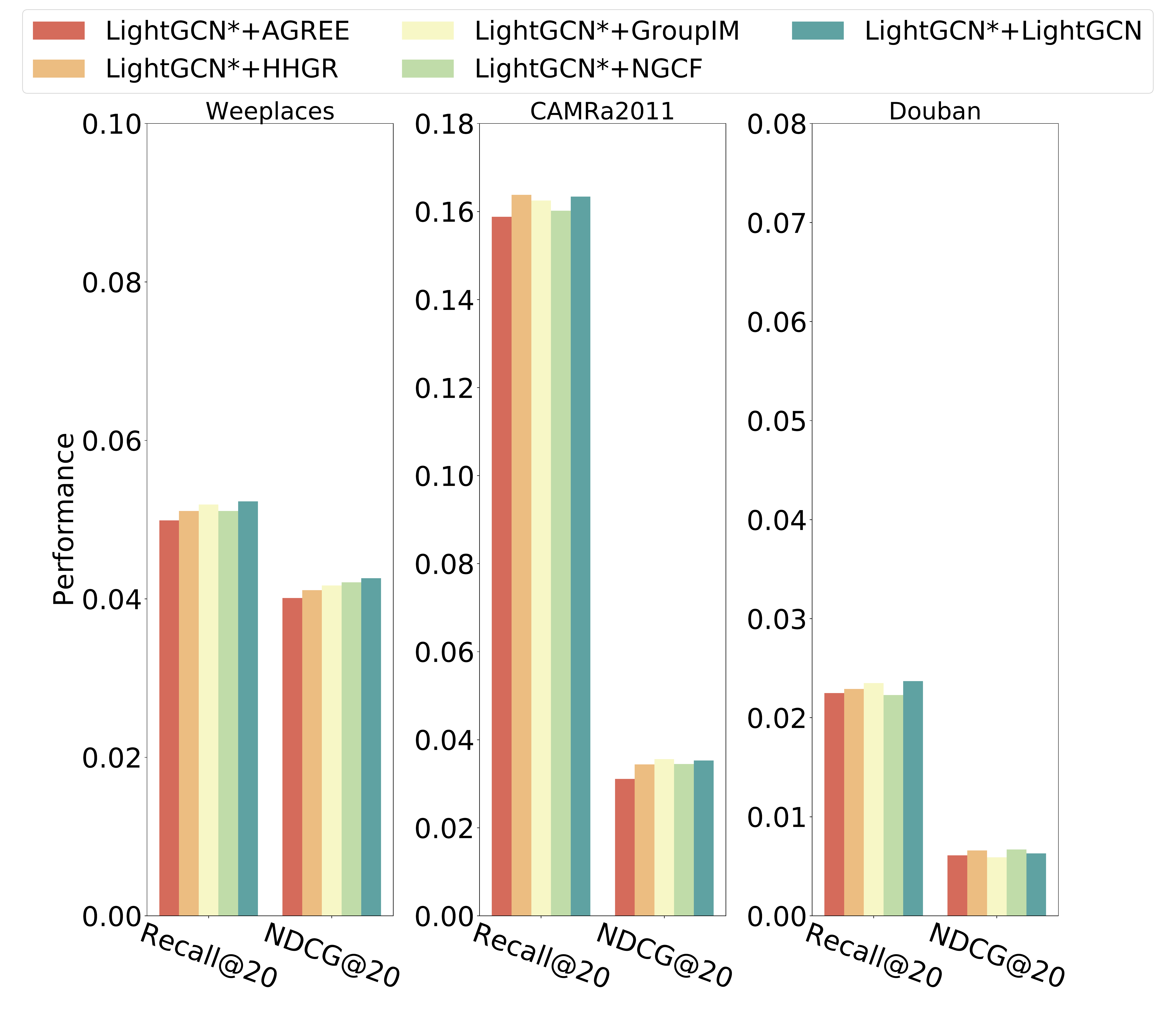}
	}
	
	\hspace{-0.4cm}
	
	\subfigure[\scriptsize  Recommendation performance 
	]{\label{subfig:gt_ngcf}
		\includegraphics[width=0.45\textwidth]{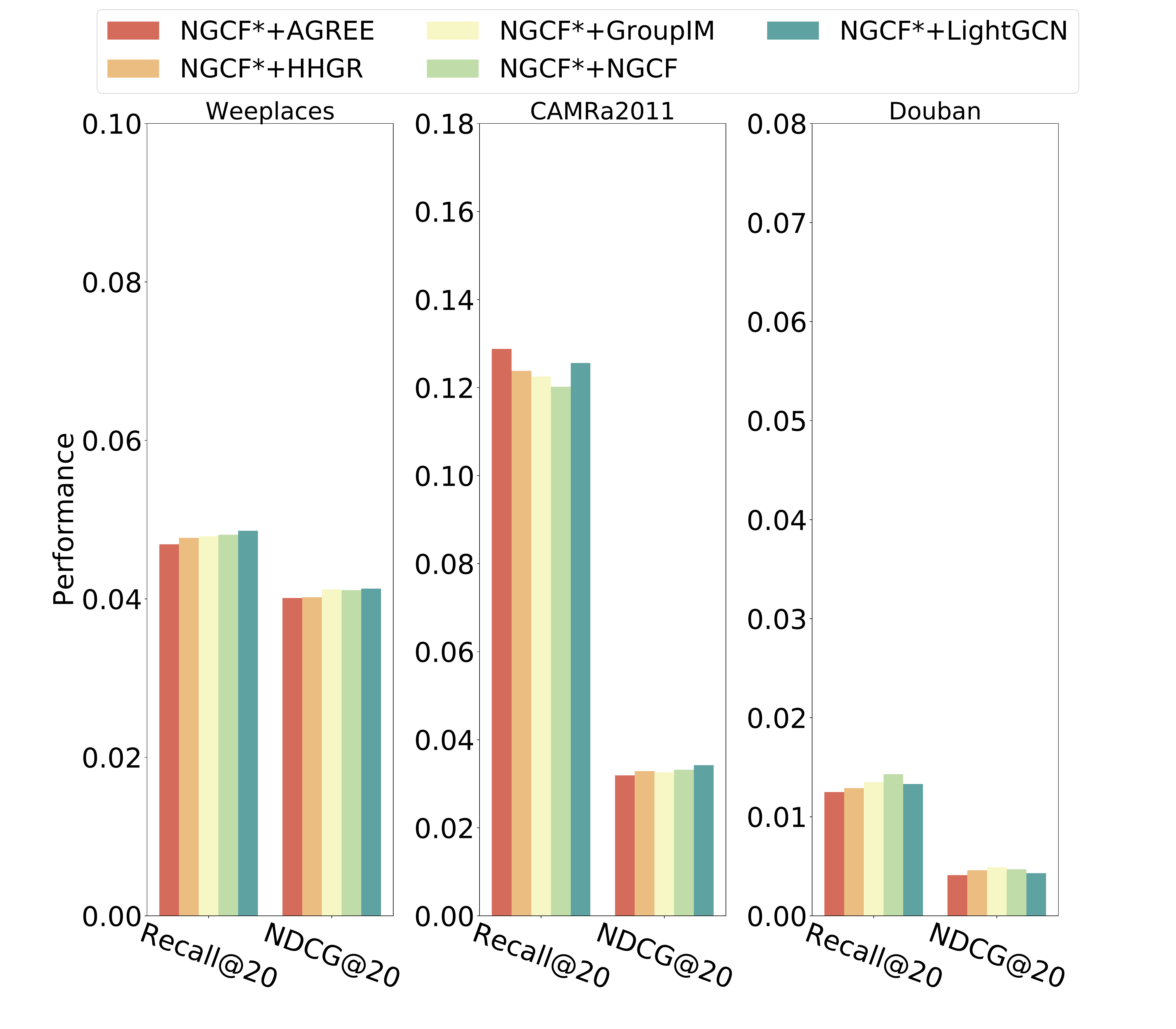}
	}
	
}

	\caption{\label{fig:sensitive_analysis} Sensitive analysis of ground-truth embeddings. }
\end{figure}

\subsubsection{Effectiveness of Ground-truth Embedding.}
\label{sec:sensitive_ground_truth}
Notably, in~\secref{sec:basic_reconstruction}, we select any group recommendation models to learn the ground-truth embeddings. 
Here we explore whether the ground-truth embeddings obtained by different models can affect the performance of \RC.
To this end, we use competitive baselines to learn the ground-truth embeddings as proposed in~\secref{sec:training_settings}, and report the performance of NGCF* and LightGCN* in Figure~\ref{fig:sensitive_analysis}. Notation NGCF*-AGREE denotes \sRC  is equipped with the ground-truth embeddings, which are obtained by AGREE. Other notations are defined in a similar way. 
The results show that the performance of \sRC is almost the same under different ground-truth embeddings, the reason is that traditional recommendation methods are able to learn accurate embeddings for the nodes with enough interactions.

\begin{table}[t]
	\caption{Recommendation performance, training time for each epoch and  convergent epochs  under the multi-task learning or  pre-training paradigms. } 
	\vspace{-10pt}
	\resizebox{0.90\textwidth}{!}{
		\begin{tabular}{l|cccc|cccc}
			\hline
			\multicolumn{1}{c|}{Dataset} & \multicolumn{4}{c|}{Weeplaces} & \multicolumn{4}{c}{CAMRa2011} \\ \hline
			\multicolumn{1}{c|}{Method}  & Recall    & NDCG  & Time &Epoch     & Recall     & NDCG   & Time & Epoch    \\ \hline\hline
			GCMC*-P                        &  0.0428           &  0.0409    &  \textbf{201.0s}    &   28 & 0.1008 & 0.0317 & \textbf{56.29s} &26                  \\
				GCMC*             & \textbf{0.0513}      & \textbf{0.0448}   &  499.6s  &  \textbf{12} & \textbf{0.1112}    & \textbf{0.0394}   & 112.3s     & \textbf{10}    \\ \hline
			NGCF*-P                                & 0.0411          &  0.0388        & \textbf{203.1s}  &  28  & 0.1182  &0.0318   & \textbf{60.2s} & 28        \\
				NGCF*            & \textbf{0.0486}    &\textbf{0.0413}   & 489.7s  & \textbf{8}  &\textbf{0.1256}     & \textbf{0.0342} & 100.8s & \textbf{8}  \\ \hline
			LightGCN*-P        &  0.0487          &   0.0386      & \textbf{189.3s}   & 28 & 0.1525&0.0327    & \textbf{50.1s}   & 28      \\
			LightGCN*                                &\textbf{0.0523}     &\textbf{0.0426}  &483.4s  & \textbf{10} &\textbf{0.1634}     &\textbf{0.0353}  & 99.3s & \textbf{6}    \\ \hline
	\end{tabular}}
	\label{tb:multi_vs_pretrain}
	\vspace{-5pt}
\end{table}

\subsubsection{ Multi-task Learning Vs Pre-training.}
\label{sec:multitask_vs_pretrain}
%The foregoing experiments have shown the effectiveness of \RC, where the main recommendation task and the SSL tasks are jointly optimized.
We report the recommendation performance under the two training paradigms as proposed in~\secref{sec:multi_task_training}.
For the pre-training paradigm, we first
pre-train the SSL task on $D_T$ and then fine-tune \sRC on $D_N$ with the recommendation task.
We use notation \RC-P to denote this training paradigm.
For the multi-task learning paradigm, we directly train the SSL task and the recommendation task on $D_T$ and $D_N$.
We report the recommendation performance, the average training time in each epoch and the average convergent epoch in Table~\ref{tb:multi_vs_pretrain}. Based on the results, we find that: (1) \RC-P performs worse than \RC, but still better than other baselines (cf. Table~\ref{tb:recommendation}). This shows the SSL task can benefit the recommendation performance. However,  jointly training the SSL task and the recommendation task is better than the pre-training \& fine-tuning paradigm, since the two tasks can enhance with each other, this finding is the same with previous findings~\cite{conf/sigir/WuWF0CLX21}. (2)  Compared with \RC-P, \sRC has faster  convergence speed.  Although \RC-P has smaller training time per epoch, it still has larger  total training time than \RC.
This verifies  the multi-task learning paradigm can  speed up the model convergence.
 %We conduct experiments ten times to compare \sRC with \RC-P, and report the average recommendation performance, the average training time in each epoch and the average convergent epoch in Table~\ref{tb:multi_vs_pretrain}. Based on the results, we find that:

\begin{figure}
	\centering
	
	\mbox{ 
		
		\subfigure[\scriptsize   Recall@20
		]{\label{subfig:layer_recall_w}
			\includegraphics[width=0.45 \textwidth]{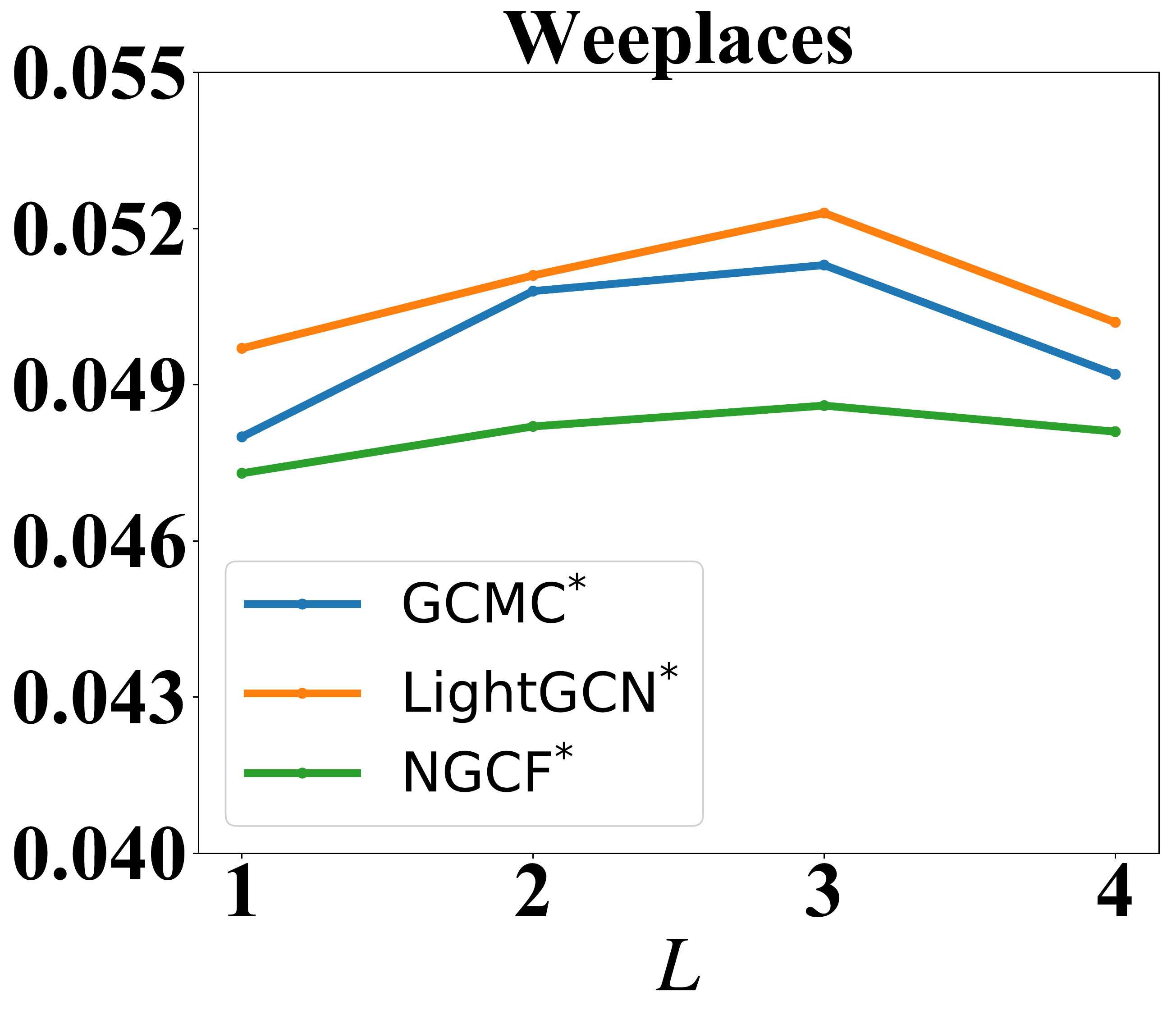}
		}

		\subfigure[\scriptsize   NDCG@20
		]{\label{subfig:layer_ndcg_w}
			\includegraphics[width=0.45\textwidth]{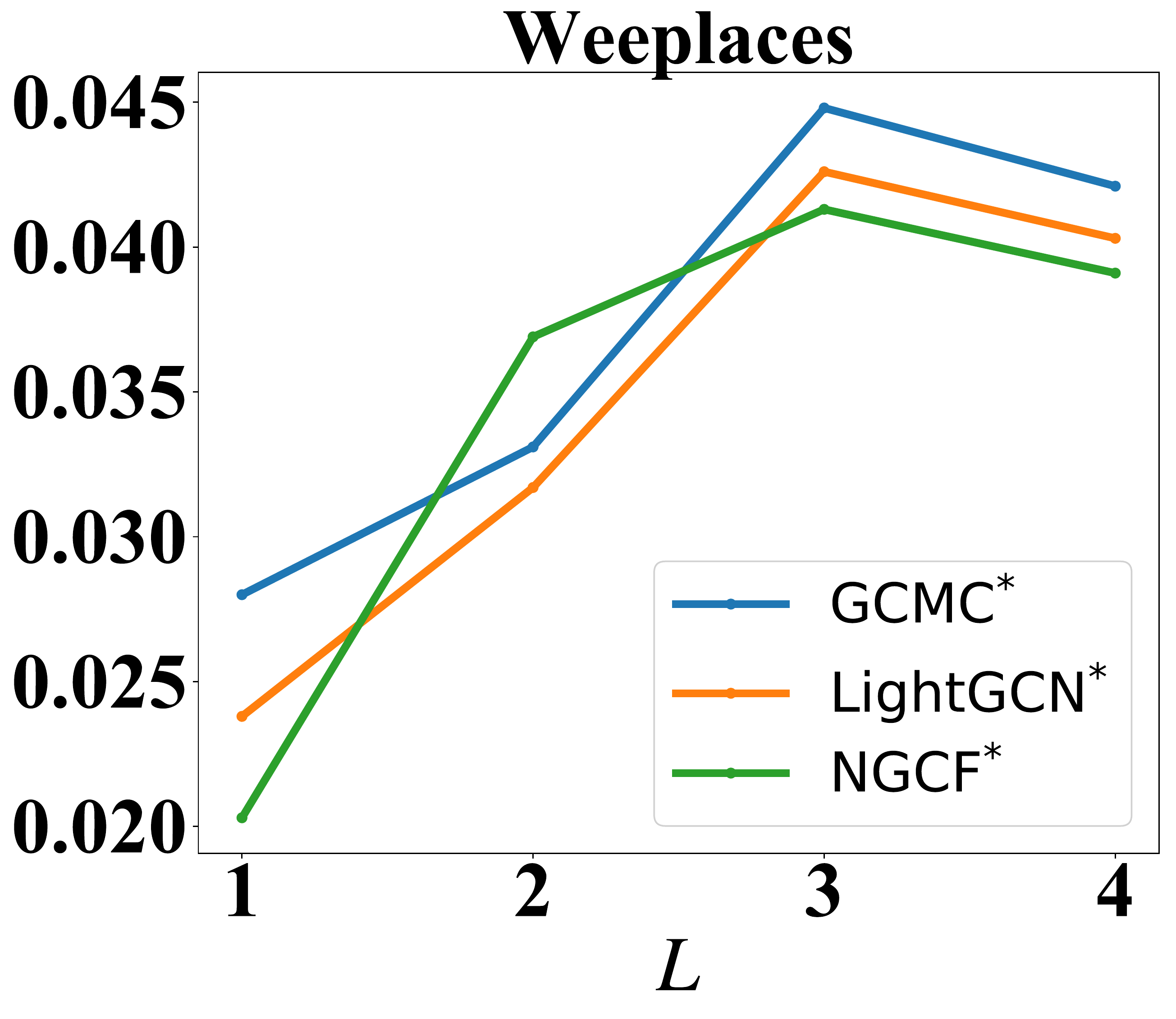}
		}
	}

	\mbox{ 
	
	\subfigure[\scriptsize   Recall@20
	]{\label{subfig:a_layer_recall_c}
		\includegraphics[width=0.45 \textwidth]{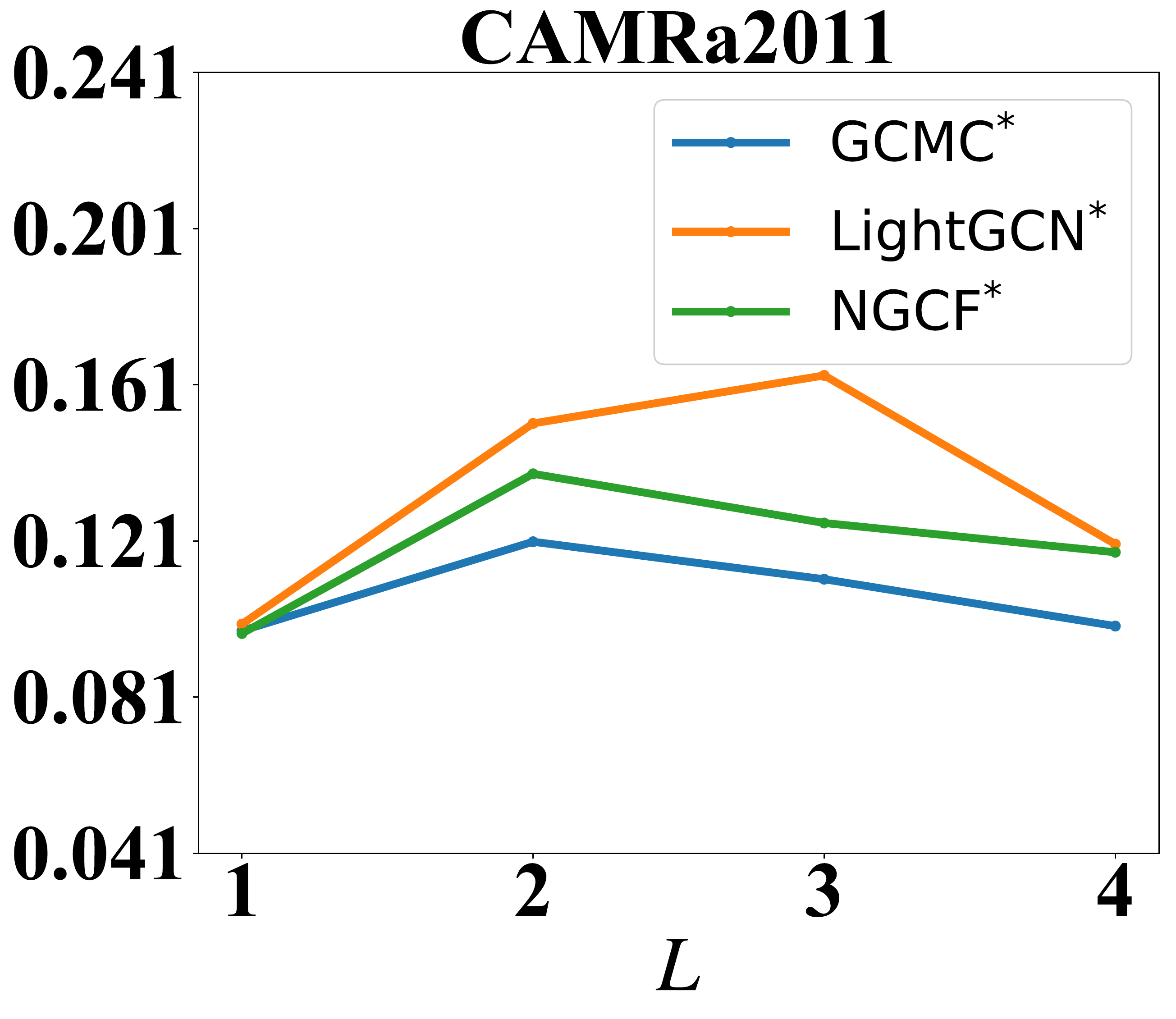}
	}
	
	\hspace{-0.2cm}
	
	\subfigure[\scriptsize  NDCG@20
	]{\label{subfig:a_layer_ndcg_c}
		\includegraphics[width=0.45\textwidth]{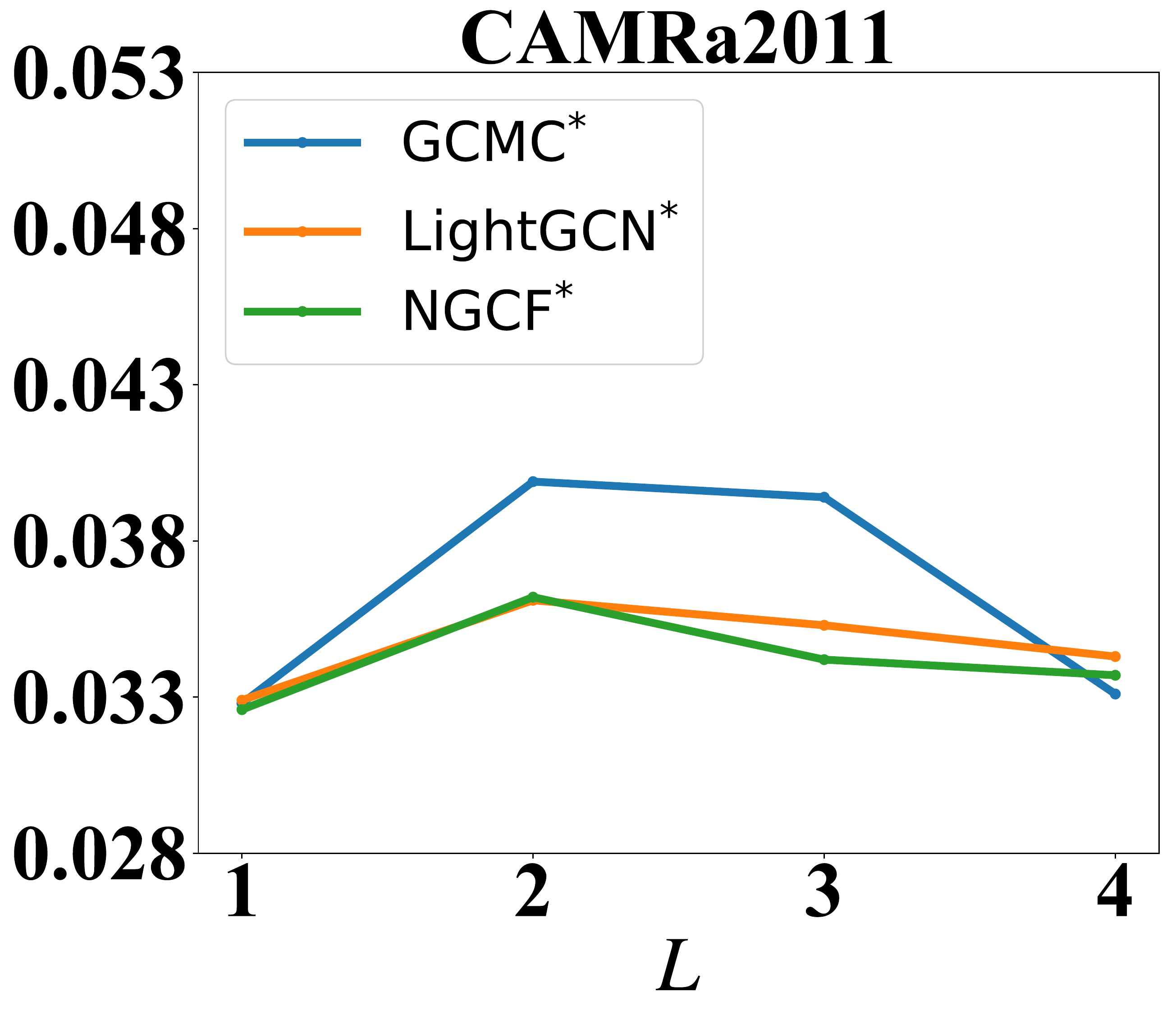}
	}
}

\mbox{	
	\subfigure[\scriptsize  Recall@20
	]{\label{subfig:a_layer_recall_d}
		\includegraphics[width=0.45 \textwidth]{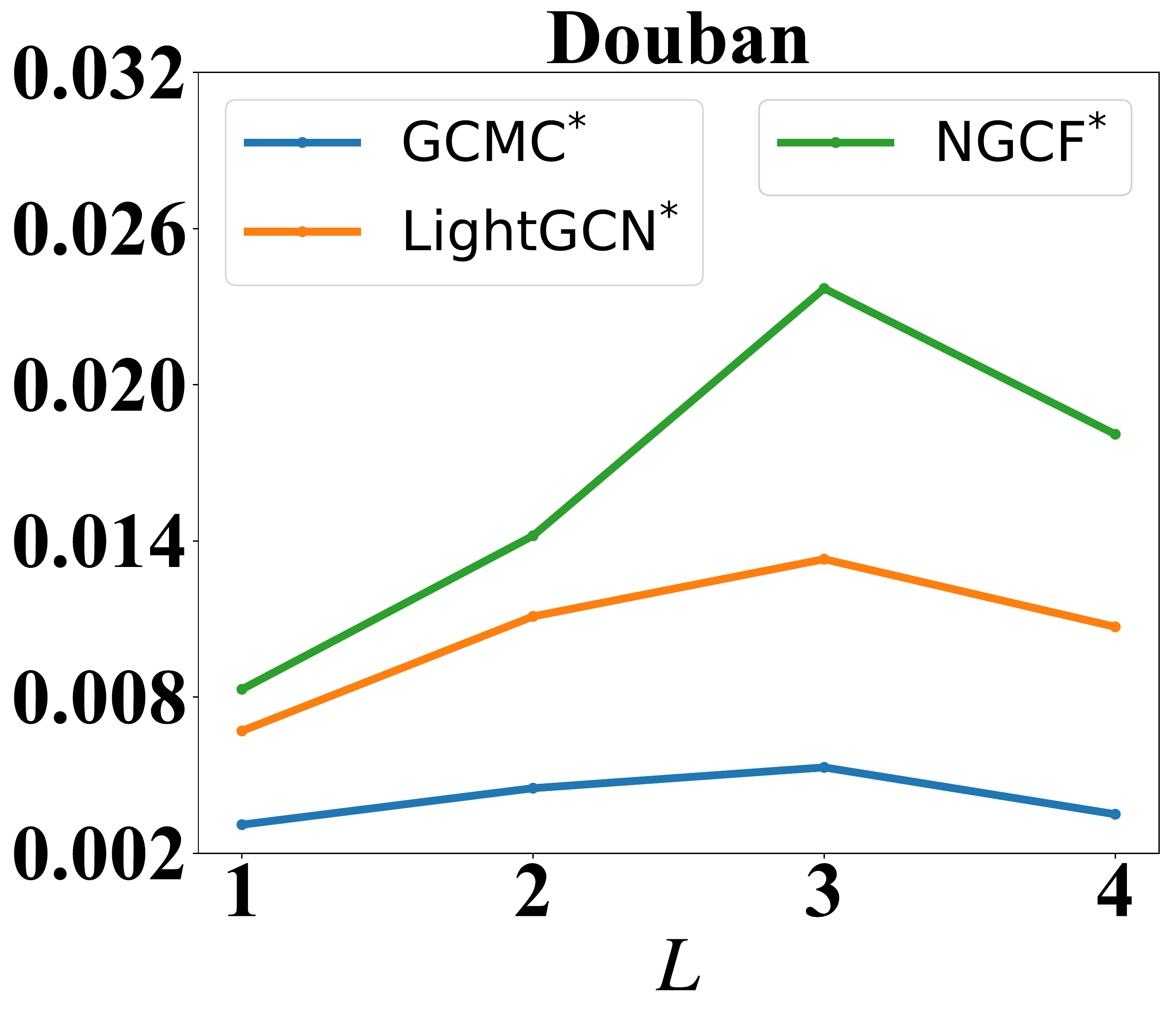}
	}
	
	\hspace{-0.2cm}
	
	\subfigure[\scriptsize  NDCG@20
	]{\label{subfig:a_layer_ndcg_d}
		\includegraphics[width=0.45\textwidth]{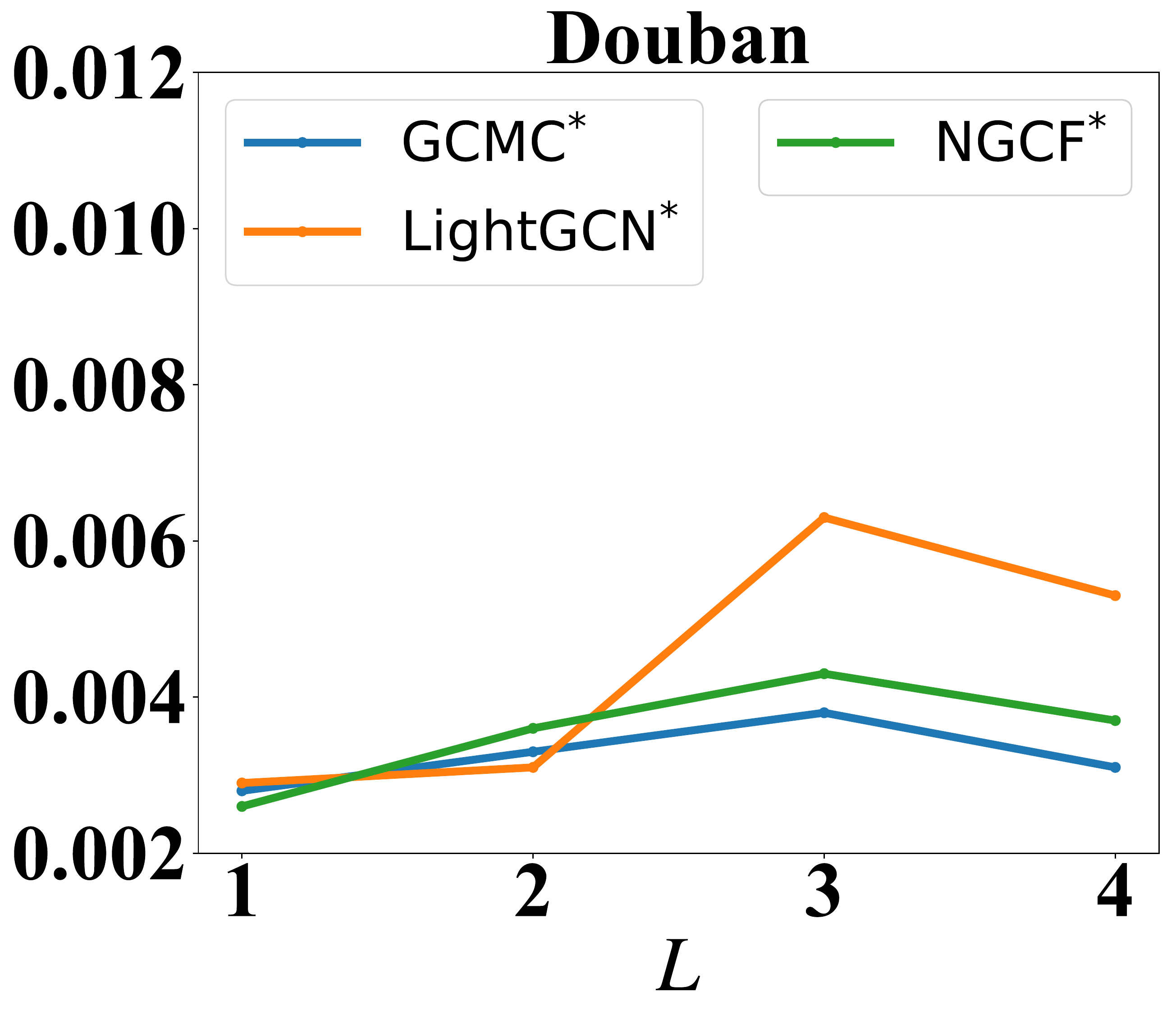}
	}
}

	\caption{ \label{fig:layer_analysis} \small{ Recommendation performance under different layer $L$ $c$\%=0.1 and $K$=5.  }}
\end{figure}

\subsubsection{Hyper-parameter analysis.}
\label{subsec:hyper_paramter}
\begin{figure}
	\centering
	
	\mbox{

		\subfigure[\scriptsize  Recall@20
		]{\label{subfig:k_recall_w}
			\includegraphics[width=0.45 \textwidth]{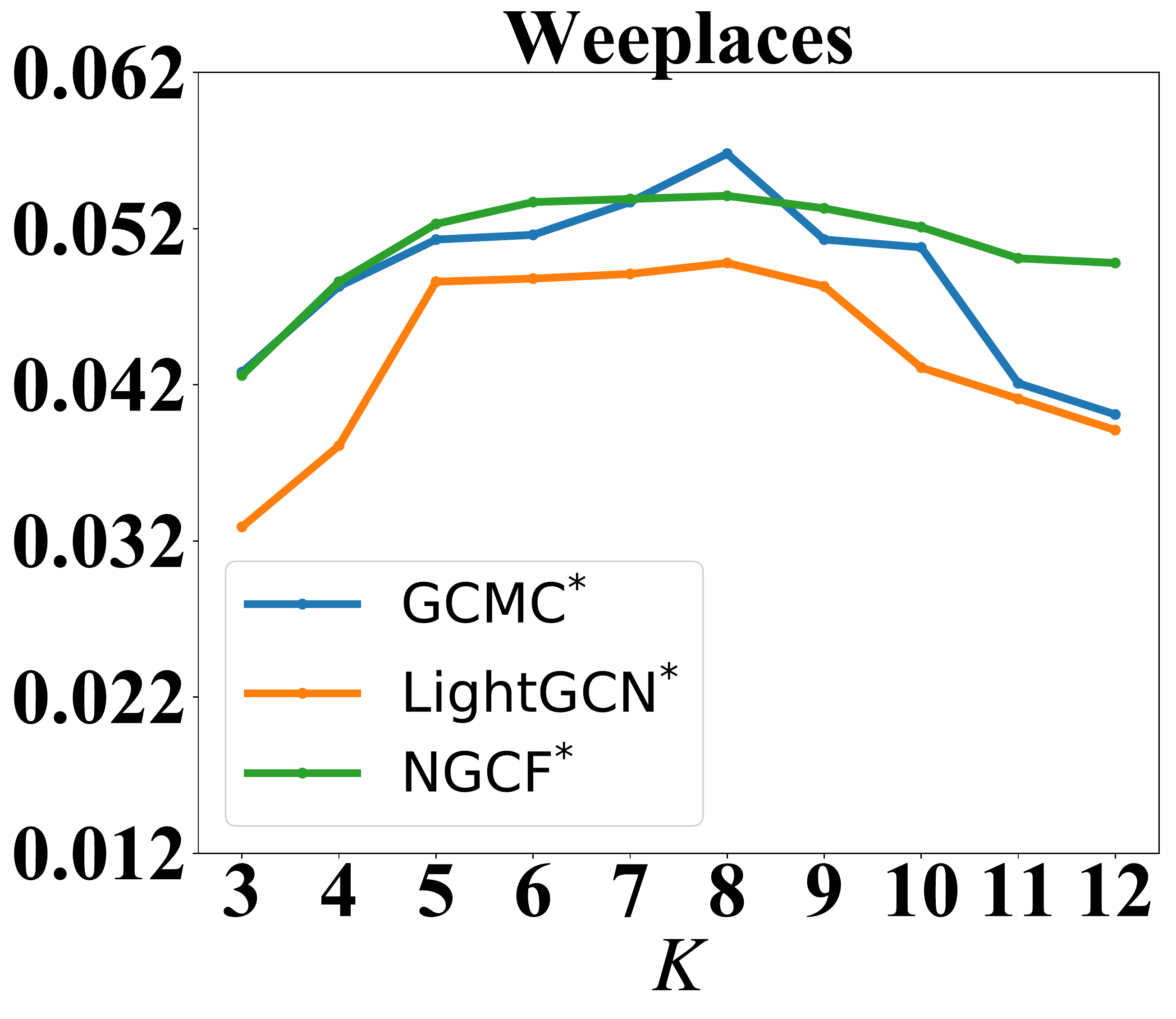}
		}
		
		\hspace{-0.2cm}
		
		\subfigure[\scriptsize  NDCG@20
		]{\label{subfig:k_ndcg_w}
			\includegraphics[width=0.45\textwidth]{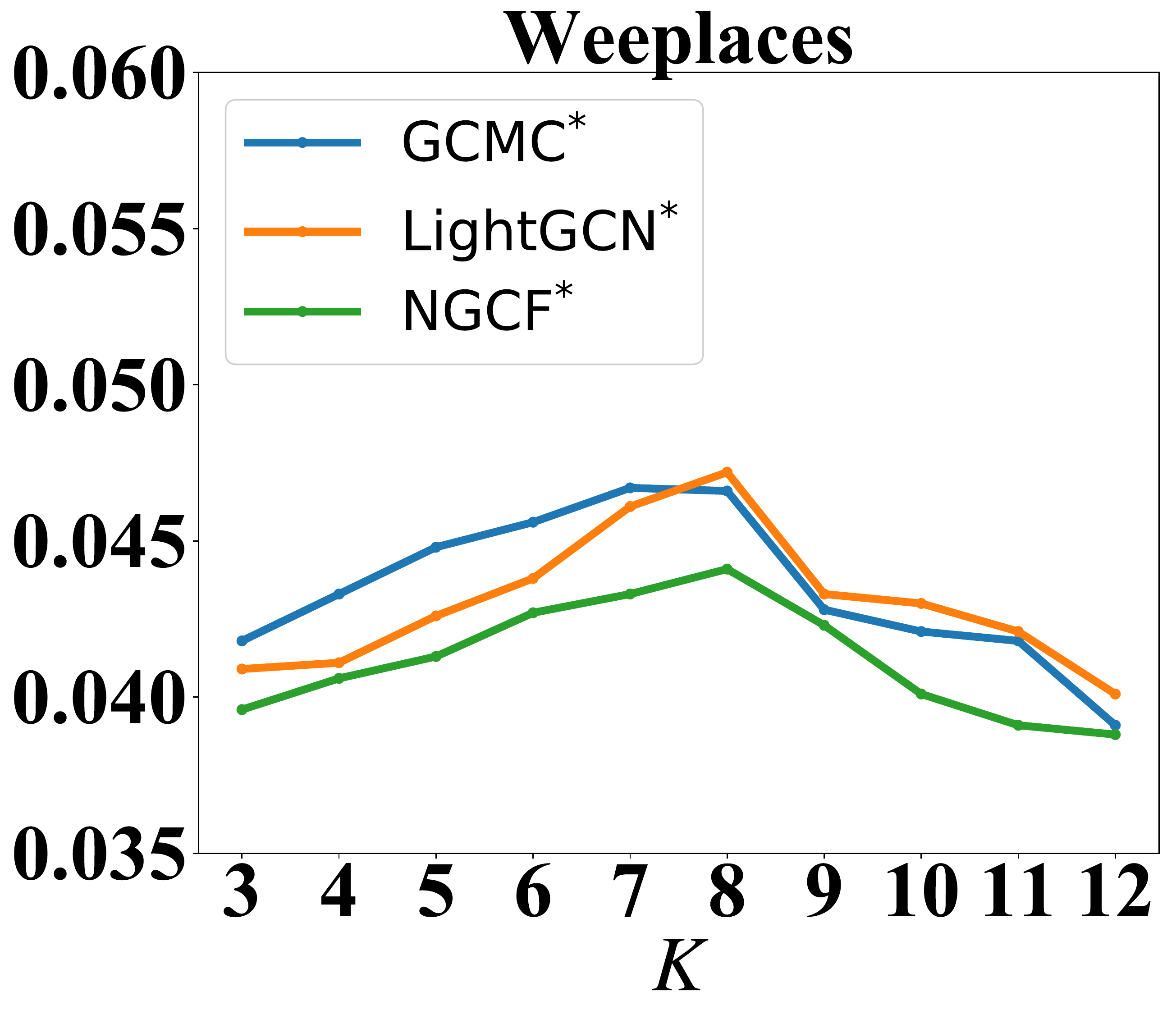}
		}
	}
	
	\mbox{ 
		
		\subfigure[\scriptsize  Recall@20
		]{\label{subfig:a_k_recall_c}
			\includegraphics[width=0.45 \textwidth]{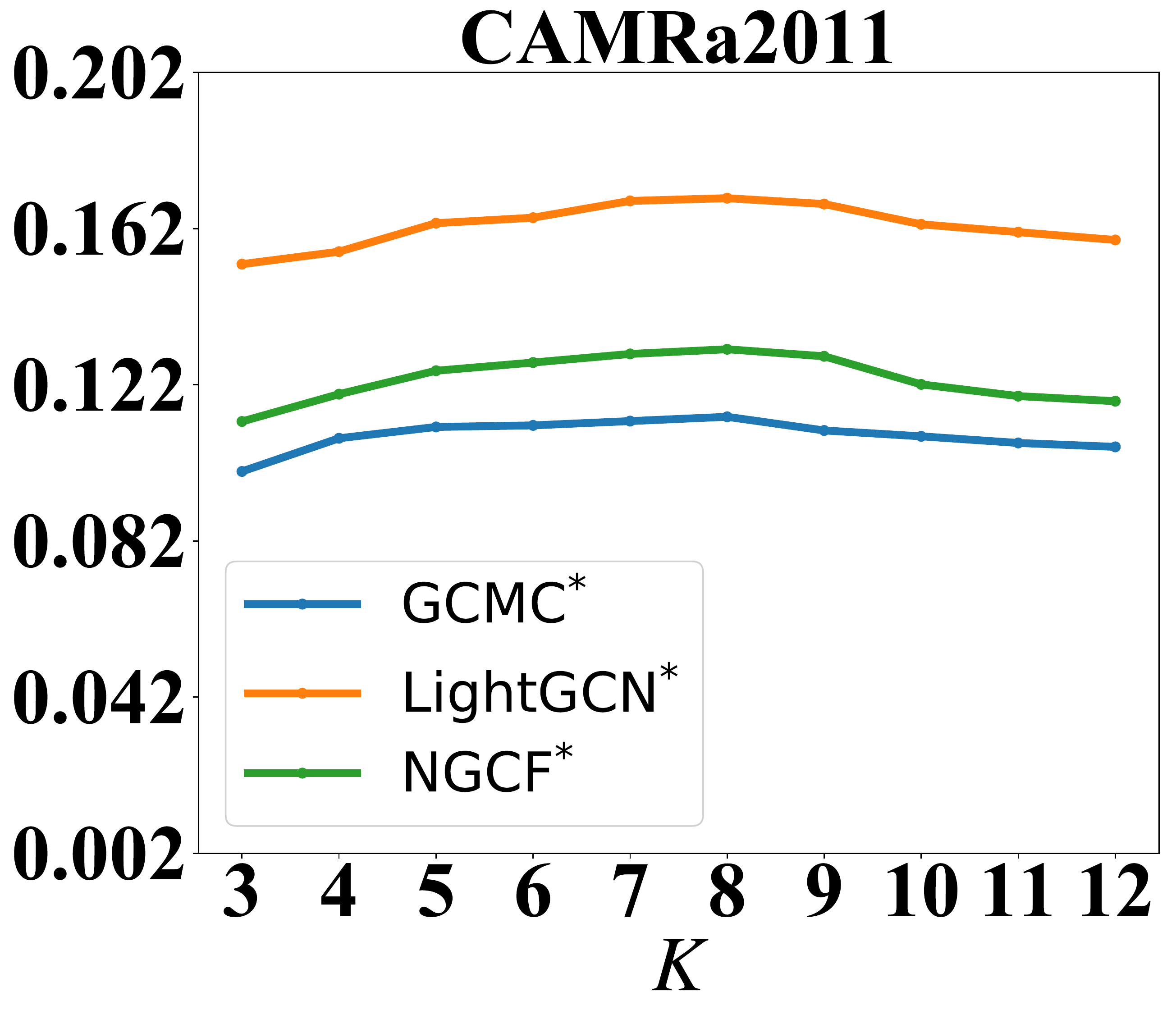}
		}
		
		\hspace{-0.2cm}
		
		\subfigure[\scriptsize  NDCG@20
		]{\label{subfig:a_k_ndcg_c}
			\includegraphics[width=0.45\textwidth]{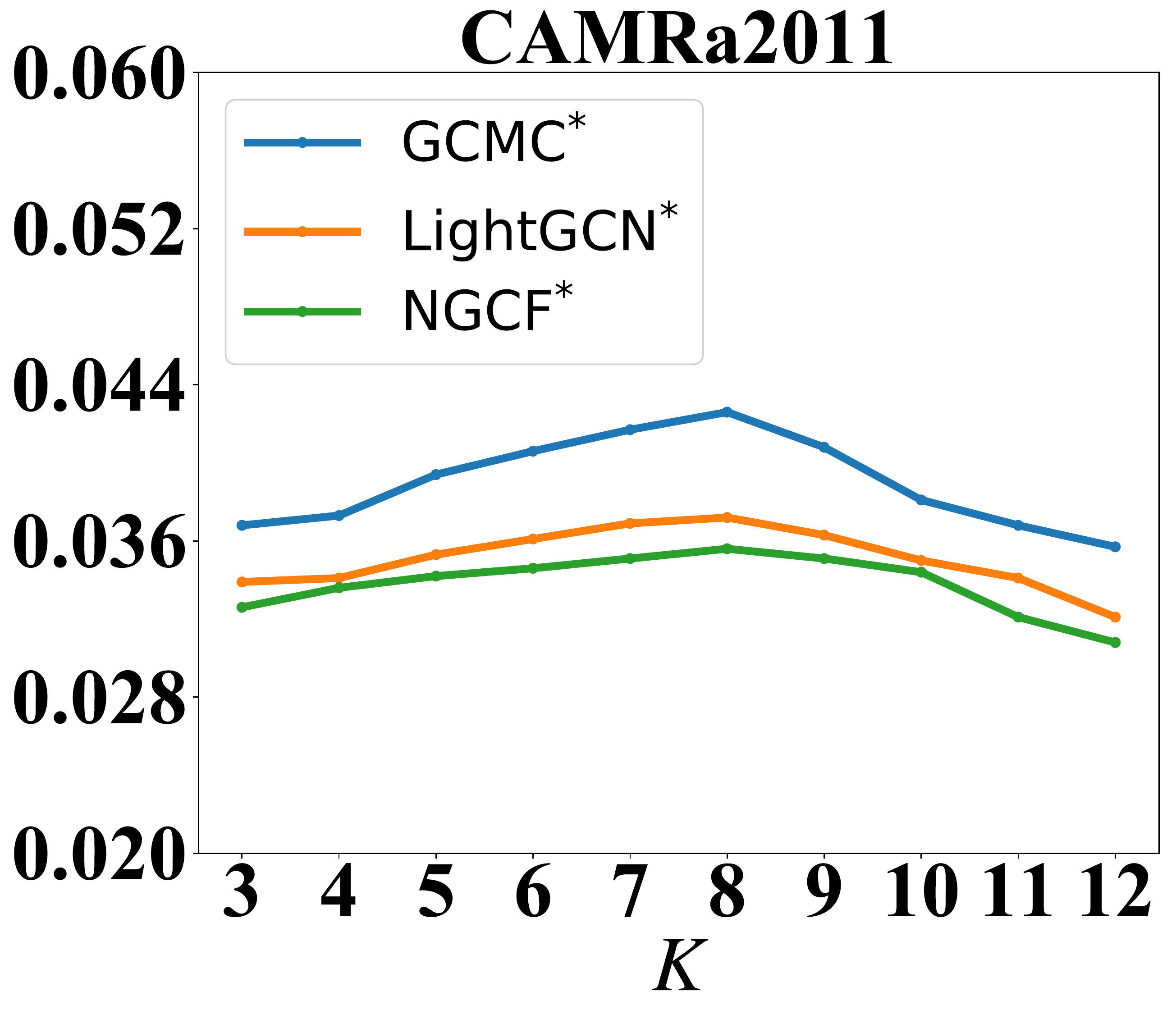}
		}
	}	
	\mbox{
		
		\subfigure[\scriptsize  Recall@20
		]{\label{subfig:a_k_recall_d}
			\includegraphics[width=0.45 \textwidth]{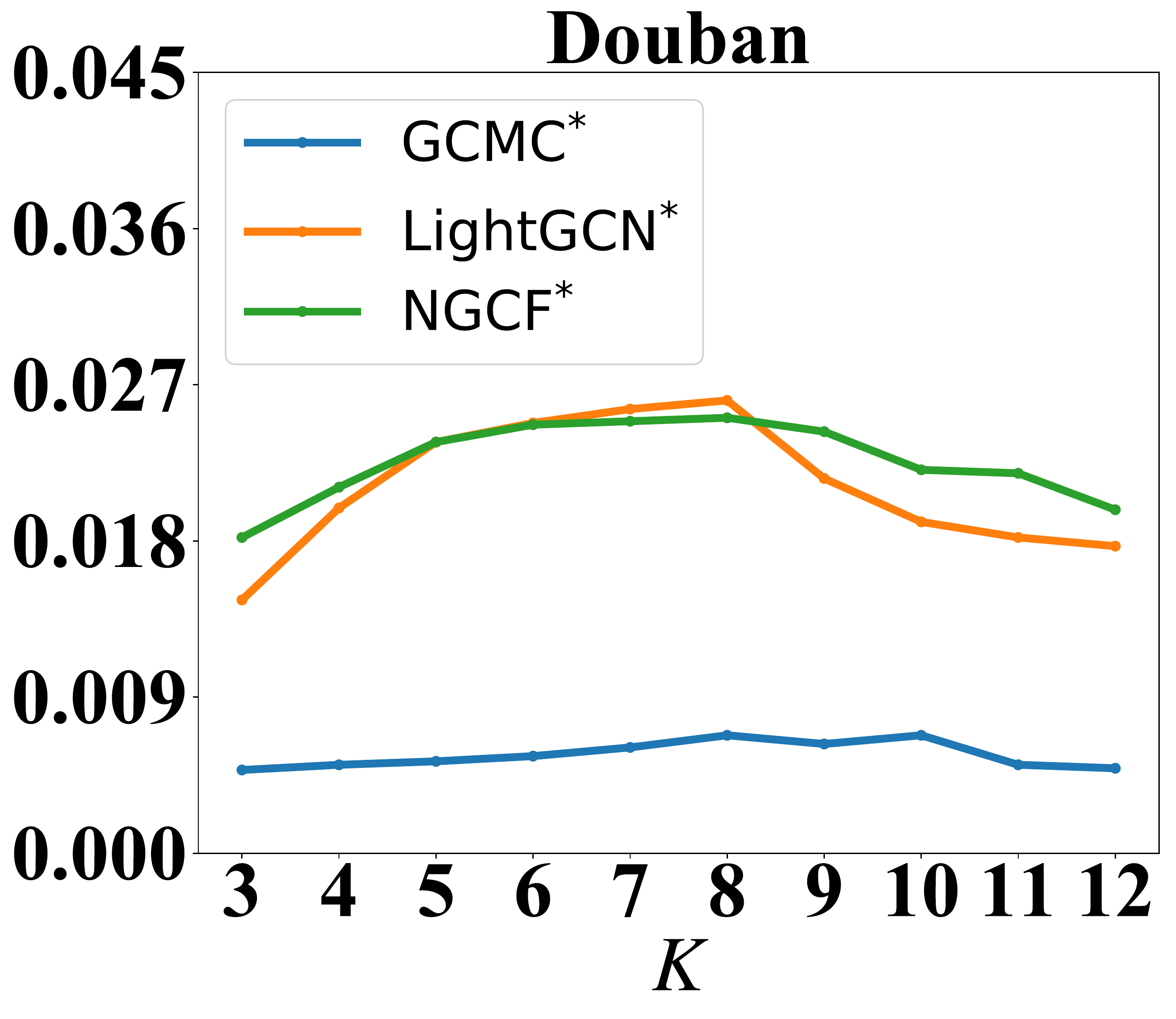}
		}
		
		\hspace{-0.2cm}
		
		\subfigure[\scriptsize  NDCG@20
		]{\label{subfig:a_k_ndcg_d}
			\includegraphics[width=0.45\textwidth]{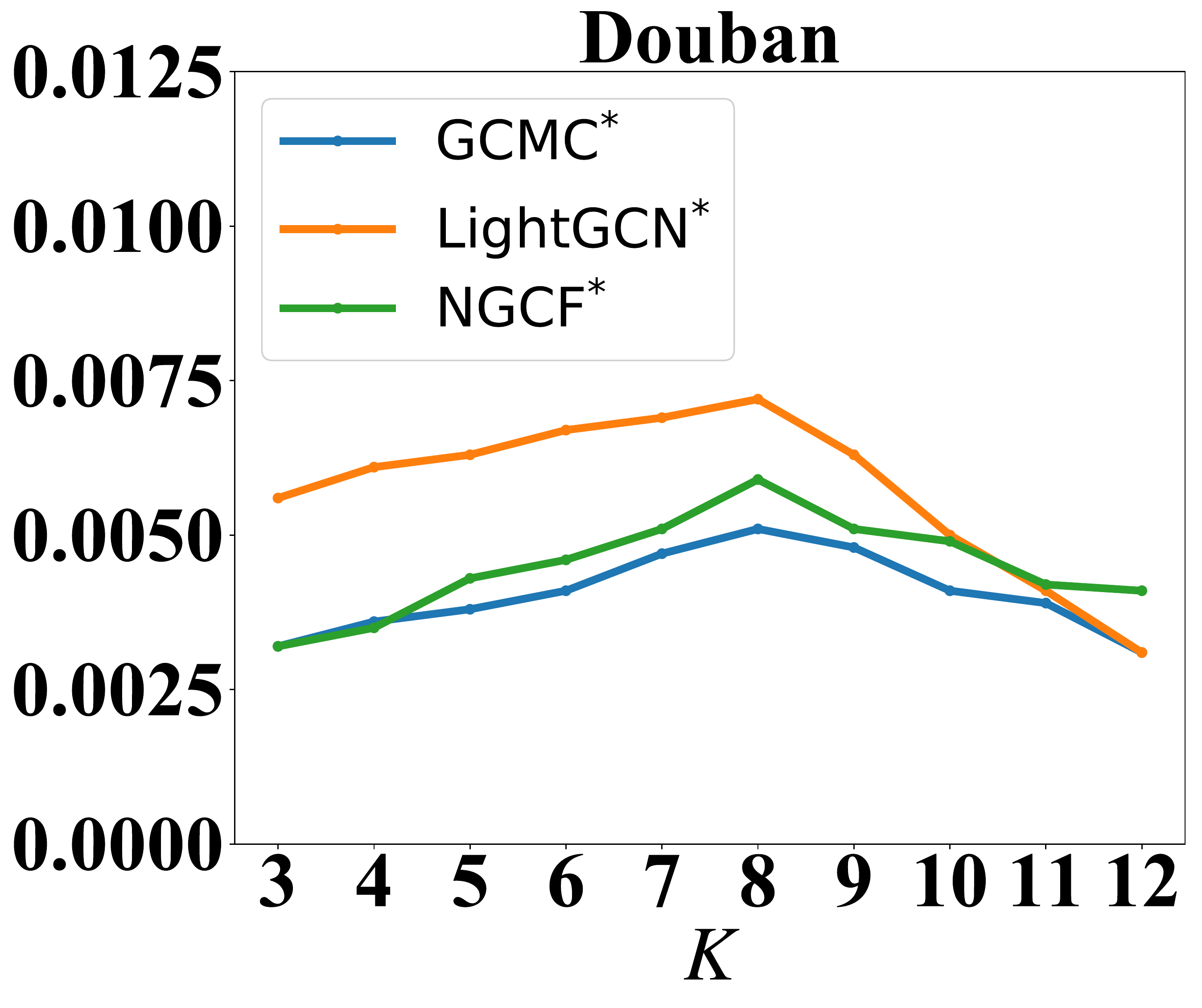}
		}
		
	}
	
	\caption{\label{fig:a_k_analysis} Recommendation performance under different neighbor size $K$.  $c$\%=0.1,  $L$=3.}
\end{figure}

\begin{figure}[t]
	\centering
	
	\mbox{
		
		\hspace{-0.5cm}
		
		\subfigure[\scriptsize  Weeplaces
		]{\label{subfig:lambda}
			\includegraphics[width=0.45 \textwidth]{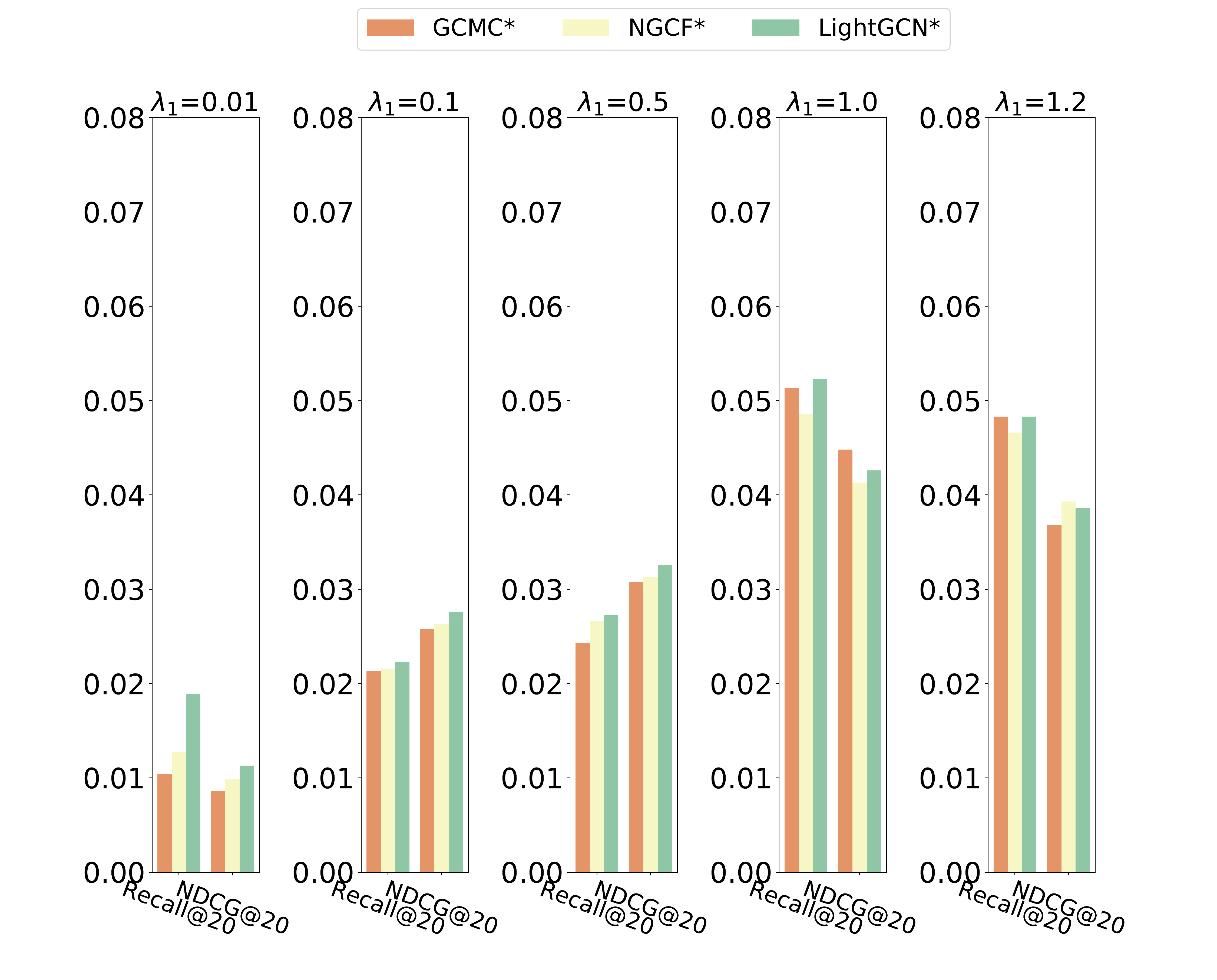}
		}
		
		\hspace{-0.5cm}
		
		\subfigure[\scriptsize  Douban
		]{\label{subfig:lambda_douban}
			\includegraphics[width=0.45 \textwidth]{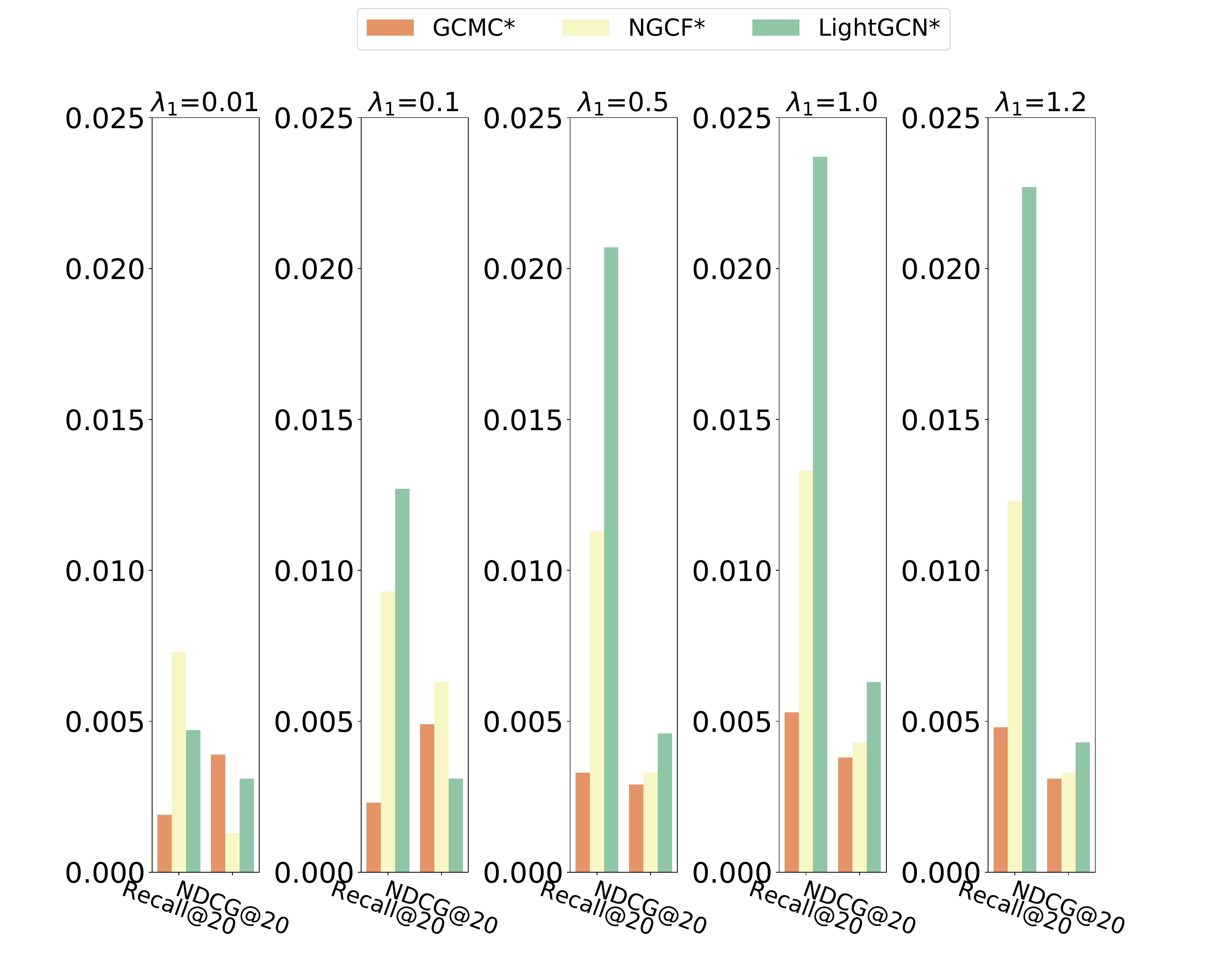}
		}
		
	}

	\caption{\label{fig:lambda_analysis} Recommendation performance under different balancing parameter $\lambda_1$.  $K$=5, $c$\%=0.1 and  $L$=3.}
\end{figure}

Here we explore whether \sRC is sensitive to the layer depth $L$, the neighbor size $K$, and the hyperparameter $\lambda1$.
We select LightGCN*, NGCF* and GCMC*,  report their performances under different layer depth $L$ in Figure~\ref{fig:layer_analysis}, under different neighbor size $K$ in Figure~\ref{fig:a_k_analysis}, and under different hyperparameter $\lambda_1$ in Figure~\ref{fig:lambda_analysis}.
The results show that:

\begin{itemize}[ leftmargin=10pt ]
	
	\item In terms of $L$, the performance increases from 1 to 3 and drops from 3 to 4.
	When $L$ equals to 3, \sRC  always has the best performance.
	This indicates using proper layer size can benefit the recommendation task.
	
	\item In terms of $K$, the performance increases from 3 to 8 and drops from 8 to 12. When $K$ is 8, \sRC can always achieve the best performance.
	This indicates incorporating proper size of neighbors can benefit the recommendation task.
	
	\item In terms of the balancing parameter $\lambda_1$, we report the recommendation performance of LightGCN*, NGCF* and GCMC*  on CAMRa2011 and Douban in Figure~\ref{fig:lambda_analysis}. The results show  the performance increases from 0.01 to 1, and drops from 1 to 1.2.
 This indicates  the auxiliary SSL tasks are  as important as the main recommendation task.

\end{itemize}

%% file: related.tex
\section{Related Work}

\subsection{Group Recommendation}
The goal of group recommendation is to recommend proper items to a group. Different from Shared-account recommendation~\cite{GCWY2022,GTCZNY2021}, where 
the members in the shared-account is closely related, the members in the group may formed at hoc. Existing methods for group recommendation can be classified into the following two categories:

\vpara{Score Aggregation Group Recommendation Strategy.}
This strategy pre-defines a scoring function to obtain the preference score of all members in a group on the target item. The scoring functions include average~\cite{BaltAverage}, least misery~\cite{pvldb/Amer-YahiaRCDY09} and maximum satisfaction~\cite{BorattoC11}.
However, due to the static recommendation process of the predefined functions, these methods easily fall into local optimal solutions.

\vpara{Profile Aggregation Group Recommendation Strategy.} This strategy  aggregates the group members' profiles and feeds the fused group profile  into individual
recommendation models. Essentially, probabilistic generative  models and deep learning based models are proposed to aggregate the group profile.
The generative model first selects group members for a target group, and then generates items based on the selected members and their associated hidden topics~\cite{cikm/LiuTYL12,sigir/YeLL12,kdd/YuanCL14}. 
The deep learning based model conducts attention mechanism to assign each user an attention
weight, which denotes the influence of group member in deciding
the group’s choice on the target item~\cite{sigir/Cao0MAYH18,tkde/CaoHMXCX21,YinW0LYZ19}.
However, both of these two methods suffer from the data sparsity issue.
Recently, researches propose GNN-based recommendation models, which incorporate high-order 
collaborative signals in the  built  graph~\cite{icde/GuoYW0HC20,sigir/Wang0RQ0LZ20,sigir/HeCZ20,YWZLZ2020} or hypergraph~\cite{yu2021self,tois/GUOYintois21group,CYLNWW2022}.
Moreover, Zhang et al.~\cite{conf/CIKM21/JMJLJH21} propose hypergraph convolution network (HHGR) with self-supervised node
dropout strategy.
However, the GNNs still can not strengthen the cold-start neighbors' embedding quality.
Motivated by recent works which leverage the SSL technique to solve the cold-start problems~\cite{YWZLZ2020,HYZLC2022,HZLC2020,CYLNWW2022} such as PT-GNN~\cite{hao2021pre} and SGL~\cite{WWFHCLX2021}, we propose the group/user/item embedding reconstruction task under the meta-learning setting, and further incorporate an embedding enhancer to improve the embedding quality.
Notably, our work is related to PT-GNN~\cite{hao2021pre}, which reconstructs the cold-start user/item embeddings for personalized recommendation. Different from PT-GNN, the group embedding reconstruction in \sRC is much more complex, as the decision process of a group is much more complicated than an individual user. Besides, the group representation not only depends on the group members' preferences, but also relies on the group-level preferences towards items and collaborative group signals. Thus, the group embedding reconstruction process is much more complicated than user/item embedding reconstruction.
Recently, Chen et al~\cite{CYLNWW2022} propose CubeRec, which uses the hypercube to model the group member decision process to enhance the 
group embedding. Although both CubeRec and \sRC is to enhance the group representation, CubeRec takes 
subspace to strengthen the group embedding, while \sRC
views the group embedding as a single point and directly leverage SSL technique to enhance the group embedding.

%% file: conclusion.tex
\section{Conclusion}
 We present a self-supervised graph learning framework for group recommendation.  In this framework, we design the user/item/group embedding reconstruction task with GNNs under the meta-learning setting, 
We further introduce an embedding enhancer  to strengthen the  GNNs' aggregation ability, which can improve the high-order cold-start neighbors' embedding quality.
Comprehensive experiments show the superiority of our proposed framework than the state-of-the-art methods.
The limitation of this work is that the proposed model  is not a general pre-training model that can be applied to new recommendation datasets.
In the future, our goal is to design a general pre-training recommendation model that can be applied to different  datasets, we hope to achieve the same effect as the natural language pre-training model BERT~\cite{DCWLT19}.
More concretely, we will dedicate to learn the structure and semantic information in the user-item-group heterogeneous graph, and transfer the learned information into new datasets.